\documentclass{pasj00}

\RelativeFigurePath{fig/}

\begin{document}
\SetRunningHead{M. Konishi et al.}{NIR Morphologies of Star-forming Galaxies at $z\sim1$}
%\Received{2010/04/12}%{yyyy/mm/dd}
%\Accepted{}%{yyyy/mm/dd}
%\Published{}%{yyyy/mm/dd}

\title{MOIRCS Deep Survey. VII: NIR Morphologies of Star-forming Galaxies at Redshift $z\sim1$}

%%%%%%%%%%%%%%%%%%%%%%%%%%%%%%%%%%%%%%%%%%%%%%%%%%%%%%%%%%%%%%%%%%%%%%%%%%%%%%%
%%% begin:list of authors
% Do NOT capitalize all letters in "textsc".

\author{Masahiro~\textsc{konishi},\altaffilmark{1,2}
        Masayuki~\textsc{akiyama},\altaffilmark{3}
        Masaru~\textsc{kajisawa},\altaffilmark{3}
        Takashi~\textsc{ichikawa},\altaffilmark{3}\\
        Ryuji~\textsc{suzuki},\altaffilmark{2}
        Chihiro~\textsc{tokoku},\altaffilmark{3}
        Yuka~\textsc{katsuno uchimoto},\altaffilmark{1}
        Tomohiro~\textsc{yoshikawa},\altaffilmark{3}\\
        Ichi~\textsc{tanaka},\altaffilmark{2}
        Masato~\textsc{onodera},\altaffilmark{4}
        Masami~\textsc{ouchi},\altaffilmark{5}
        Koji~\textsc{omata},\altaffilmark{2}
        Tetsuo~\textsc{nishimura},\altaffilmark{2}\\
        and
        Toru~\textsc{yamada}\altaffilmark{3}
        }
\altaffiltext{1}{Institute of Astronomy, University of Tokyo, Mitaka, Tokyo 181-0015, Japan}
\email{konishi@ioa.s.u-tokyo.ac.jp}
\altaffiltext{2}{Subaru Telescope, National Astronomical Observatory of Japan, 650 North A'ohoku Place, Hilo, HI 96720, USA}
\altaffiltext{3}{Astronomical Institute, Tohoku University, Aramaki, Aoba, Sendai 980-8578, Japan}
\altaffiltext{4}{CEA-Saclay, DSM/DAPNIA/Service d'Astrophysique, 91191 Gif-sur-Yvette Cedex, France}
\altaffiltext{5}{Observatories of the Carnegie Institution of Washington, 813 Santa Barbara Street, Pasadena, CA 91101, USA}

%%%%%%%%%%%%%%%%%%%%%%%%%%%%%%%%%%%%%%%%%%%%%%%%%%%%%%%%%%%%%%%%%%%%%%%%%%%%%%%
%% `\KeyWords{}' always has to be placed before `\maketitle'.
\KeyWords{galaxies: spiral  --- galaxies: starburst  --- galaxies: structure --- infrared: galaxies} %Do NOT move this preamble from here!

\maketitle

%%%%%%%%%%%%%%%%%%%%%%%%%%%%%%%%%%%%%%%%%%%%%%%%%%%%%%%%%%%%%%%%%%%%%%%%%%%%%%%
\begin{abstract}
We investigate rest-frame near-infrared (NIR) morphologies of a sample of 139 galaxies with $M_{\mathrm{s}}$ $\geq 1\times10^{10}$ $M_{\solar}$ at $z$=0.8--1.2 in the GOODS-North field using our deep NIR imaging data (MOIRCS Deep Survey, MODS).
We focus on Luminous Infrared Galaxies (LIRGs), which dominate high star formation rate (SFR) density at $z\sim1$, in the sample identified by cross-correlating with the \textit{Spitzer}/MIPS 24$\mu$m source catalog.
We perform two-dimensional light profile fitting of the $z\sim1$ galaxies in the $K_{\mathrm{s}}$-band (rest-frame $J$-band) with a single component $\rm{S\acute{e}rsic}$ model.
We find that at $z\sim1$, $\sim$90\% of LIRGs have low $\rm{S\acute{e}rsic}$ indices ($n<2.5$, similar to disk-like galaxies) in the $K_{\mathrm{s}}$-band, and those disk-like LIRGs consist of $\sim$60\% of the whole disk-like sample above $M_{\mathrm{s}}$ $\geq$ $3\times10^{10}$ $M_{\solar}$.
The $z\sim1$ disk-like LIRGs are comparable or $\sim20$\% small at a maximum in size compared to local disk-like galaxies in the same stellar mass range.
If we examine rest-frame UV--optical morphologies using the \textit{HST}/ACS images, the rest-frame $B$-band sizes of the $z\sim1$ disk-like galaxies are comparable to those of the local disk-like galaxies as reported by previous studies on size evolution of disk-like galaxies in the rest-frame optical band.
Measuring color gradients (galaxy sizes as a function of wavelength) of the $z\sim1$ and local disk-like galaxies, we find that the $z\sim1$ disk-like galaxies have 3--5 times steeper color gradient than the local ones.
Our results indicate that (i) more than a half of relatively massive disk-like galaxies at $z\sim1$ are in violent star formation epochs observed as LIRGs, and also (ii) most of those LIRGs are constructing their fundamental disk structure vigorously.
The high SFR density in the universe at $z\sim1$ may be dominated by such star formation in disk region in massive galaxies.
\end{abstract}

%%%%%%%%%%%%%%%%%%%%%%%%%%%%%%%%%%%%%%%%%%%%%%%%%%%%%%%%%%%%%%%%%%%%%%%%%%%%%%%
\section{Introduction}

It has been well known that the star formation rate (SFR) density of the universe (cosmic SFR density) increases by an order of magnitude from the present to $z\sim1$ (\cite{lilly95}; \cite{madau96}; \cite{hogg98}; \cite{flores99}; \cite{haarsma00}; \cite{hopkins04}; \cite{hopkins06}).
What kind of galaxies are contributing to such a large cosmic SFR density at $z\sim1$ is still an open question.

In recent years with the \textit{Spitzer Space Telescope}, many studies focusing on understanding the properties of distant star-forming galaxies have been conducted.
The cosmic SFR densities at high redshifts are dominated by infrared bright galaxies, especially Luminous Infrared (IR) Galaxies (LIRGs) which have IR (8--1000 $\mu$m) luminosities, $L_{\mathrm{IR}}$, of $10^{11}$--$10^{12}$ $L_{\solar}$ (\cite{flores99}; \cite{takeuchi05}; \cite{lefloch05}; \cite{caputi07}; \cite{perez08a}; \cite{magnelli09}; see \cite{sanders96} for a review on LIRGs).
Their luminous IR emission is thermal dust emission in the mid-IR (MIR) to far-IR (FIR) wavelength caused by absorbing UV photons from intensive star formation and/or luminous active galactic nuclei (AGNs).
IR luminosity of $10^{11}$ $L_{\solar}$ corresponds to the SFR of $\sim$ 15 $M_{\solar}$ yr$^{-1}$ (\cite{kennicutt98}), which is several to ten times larger than that of local normal galaxies (\cite{brinchmann04}).
Although in the local universe LIRGs are rare and have only $\sim5$\% contribution to the IR luminosity density, the number density of LIRGs increases with redshift, and 70\% of the IR luminosity density is in LIRGs at $z\sim0.7$ (\cite{lefloch05}). 
It is important to understand triggering processes of the LIRGs to unveil the physical cause of the evolution of the cosmic SFR densities.

In the local universe, most of LIRGs show irregular morphologies, indicating galaxy-galaxy interaction (\cite{sanders96}; \cite{sanders04}).
Even among the LIRGs which have spiral morphology apparently, a significant fraction of them shows bar structure (\cite{wang06}).
Such morphological properties indicate that gas is pushed into the nuclear region and nuclear starburst is triggered.

On the contrary, at high redshifts ($0.7 \lesssim z \lesssim 1$), the \textit{Hubble Space Telescope} (\textit{HST}) imagings show that most of LIRGs have spiral (or late-type) morphology without any clear sign of merging and/or interaction (\cite{zheng04}; \cite{bell05}; \cite{mel05}; \cite{lotz08}).
Those galaxies also have no bar structure (\cite{zheng05}).
Moreover, in the last couple of years, some studies dedicated on MIR spectra of such distant LIRGs found their spectral shape indicates cool dust temperature similar to local spiral galaxies with lower IR luminosity ($L_{\mathrm{IR}}$ $<10^{10.5}$ $L_{\solar}$) (\cite{zheng07}; \cite{symeonidis09}; \cite{seymour10}).
More recently, \citet{takagi10} reported that most of polycyclic aromatic hydrocarbon (PAH)-selected galaxies with $L_{\mathrm{IR}}>10^{11}L_{\mathrm{\solar}}$ at $z\sim1$ show PAH-to-total IR luminosity ratio similar to that of less luminous starburst galaxies using an AKARI multi-wavelength MIR photometry.
Those distant LIRGs show similar properties to local massive spiral galaxies except for their high IR luminosities.

Previous morphological studies of LIRGs at high redshifts were conducted in the rest-frame optical range.
However, the high IR luminosity of LIRGs indicates the presence of a large amount of dust, and then morphological analysis in Near-IR (NIR) range less affected by dust-extinction is necessary to examine existence of dust-obscured structures such as nuclear starburst seen in local LIRGs or bars invisible in optical.
NIR wavelength is also suitable to reveal the distribution of old stellar populations which dominate stellar mass of galaxies.
As a pioneering study in NIR, \citet{mel08} carried out targeted observations of 15 LIRGs at $0.4 \lesssim z \lesssim 1.2$ with Adaptive Optics (AO) in the $K$-band, and evaluated their morphologies visually.
It was found that two thirds of the LIRGs look disky apparently while major mergers are $\sim30$\% at a maximum. 
They also found that only one LIRG has a clear bar structure.
The distant LIRGs are different from local LIRGs in morphology, and star formation in the distant LIRGs seems to be smoothly distributed in their disk.

However, the morphological study by \citet{mel08} is based on the visual classification.
Further NIR morphological studies with quantitative evaluation are necessary.
In addition, although they observed normal galaxies (non-LIRGs) at the similar redshifts, 
it is still crucial to locate LIRGs in field galaxies in the same redshift range.

In this paper, we construct a stellar mass- and volume-limited, NIR morphological catalog of galaxies at $z\sim1$ independent of IR luminosity using deep $K_{\mathrm{s}}$-band data, and locate LIRGs among $z\sim1$ galaxies.
We evaluate the galaxy morphology quantitatively using two-dimensional light profile in both $K_{\mathrm{s}}$- (rest-frame $J$-) and optical (rest-frame $U$- and $V$-) bands. 
Comparing the rest-frame NIR and UV-to-optical morphologies, we discuss distributions of old stellar population and currently star-forming region in the high redshift LIRGs and non-LIRG galaxies.
Throughout this paper, we use a $\Lambda$-CDM cosmology with $\Omega_{m}$ = 0.3, $\Omega_{\Lambda}$ = 0.7, and $H_{0}$ = $70$ km s$^{-1}$ Mpc$^{-1}$. In this cosmology, 1 arcsec corresponds to 8.01 kpc at $z=1$.
We refer \textit{HST}/Advanced Camera for Surveys (ACS) filters F435W, F606W, F775W, and F850LP as $B_{435}$, $V_{606}$, $i_{775}$, and $z_{850}$-bands, respectively.

%%%%%%%%%%%%%%%%%%%%%%%%%%%%%%%%%%%%%%%%%%%%%%%%%%%%%%%%%%%%%%%%%%%%%%%%%%%%%%%
\section{Sample of z $\sim$ 1 Galaxies}

%%%%%%%%%%%%%%%%%%%%%%%%%%%%%%%%%%%%%%%%%%%%%%%%%%%%%%%%%%%%%%%%%%%%%%%%%%%%%%%
\subsection{$K_{\mathrm{s}}$-selected stellar mass catalog from MODS Deep Data}
\label{sect:mods}

\begin{figure*}
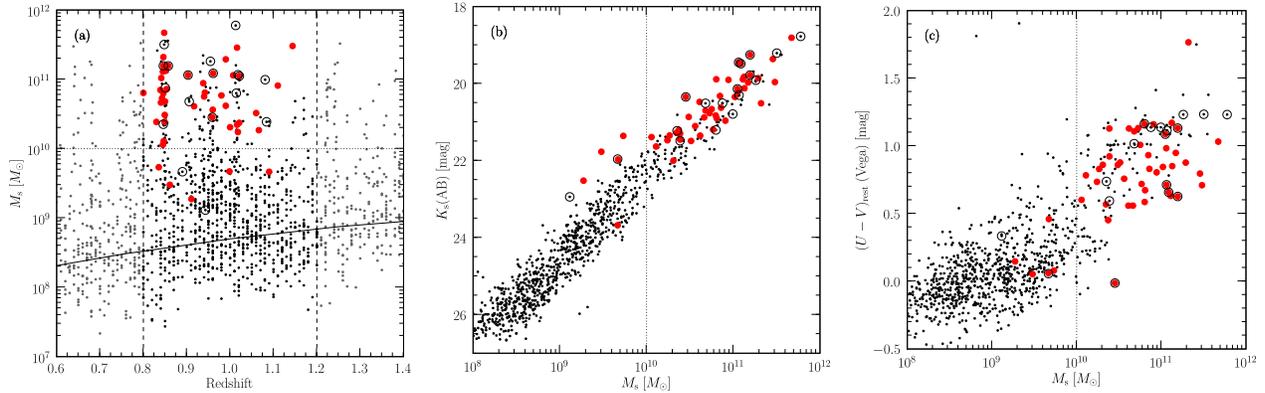

  \begin{center}
    \FigureFile(55mm,40mm){figure01a.eps}
    \FigureFile(55mm,40mm){figure01b.eps}
    \FigureFile(55mm,40mm){figure01c.eps}
  \end{center}
  \caption{Galaxies around $z=1$ in the MODS deep region. (a) Stellar mass distribution of galaxies at $z=$ 0.6--1.4. The \textit{dashed} lines separate our $z\sim1$ ($z$ = 0.8--1.2) galaxies from the $K_{\mathrm{s}}$-selected sample.  The \textit{solid} line indicates the stellar mass limit corresponding to $K_{\mathrm{s}}$(Vega) $=24$ mag (\cite{kajisawa09}, K09). The \textit{dotted} line indicates a stellar mass limit for our morphological sample selection ($M_{\mathrm{s}}$ $=1 \times 10^{10}$ $M_{\solar}$) defined in Section \ref{sect:result_nK_ReK}. Large \textit{filled} circles (red) represent MIPS 24 $\mu$m-detected ($f_{24}$ $\geq$ 80 $\mu$Jy) sources. \textit{Chandra} X-ray-detected sources are marked with large \textit{open} circle. (b) Stellar masses and $K_{\mathrm{s}}$ magnitudes of galaxies at $z=$ 0.8--1.2. (c) Stellar masses and rest-frame $U-V$ colors of galaxies at $z=$ 0.8--1.2. In all panels, the uncertainties of the stellar mass estimates are less than 0.1 and 0.05 dex for galaxies with $M_{\mathrm{s}}$ $=10^{10}-10^{11}$ $M_{\solar}$ and with $M_{\mathrm{s}}$ $> 10^{11}$ $M_{\solar}$, respectively.}
  \label{fig:MODSsample}
\end{figure*}

We construct a stellar mass-limited $z\sim1$ galaxy sample using the \textit{deep} catalog of Kajisawa et al. (2009, hereafter K09).
The catalog contains 3203 $K_{\mathrm{s}}$(Vega) $<24$ mag galaxies in 28 arcmin$^{2}$ from MOIRCS (Multi-Object InfraRed Camera and Spectrograph, \cite{suzuki08}) Deep Survey (MODS) ultra-deep image obtained with MOIRCS attached to the Subaru Telescope (\cite{kajisawa06}).
It covers a part of the Great Observatories Origins Deep Survey North (GOODS-N, \cite{giava04}) field including Hubble Deep Field North region (HDF-N) .
Utilizing publicly available multi-wavelength ($U$ $B_{435}$ $V_{606}$ $i_{775}$ $z_{850}$ $J$ $H$ $K_{\mathrm{s}}$, $3.6\mu$m, $4.5\mu$m, and $5.8\mu$m) data and spectroscopic redshift ($z_{\mathrm{spec}}$) catalogs in the GOODS-N field, K09 constructed a catalog of redshift and stellar mass for the $K_{\mathrm{s}}$-selected sample.
The redshifts and stellar masses were estimated by fitting model spectral energy distributions (SEDs) obtained from several population synthesis models to the SEDs of the sample.
The Salpeter Initial Mass Function (IMF) in stellar mass range of $0.1-100$ $M_{\solar}$ (\cite{salpeter55}) is assumed in the SED fittings.
In this paper, we adopt the photometric redshifts ($z_{\mathrm{phot}}$) and stellar masses ($M_{\mathrm{s}}$) obtained with \textsc{Galaxev} population synthesis model (\cite{bc03}). 
The comparison between the photometric and spectroscopic redshifts of the galaxies which have the spectroscopic data shows that the photometric redshift estimates have a good accuracy with median and standard deviation of $\delta z \equiv$ ($z_{\mathrm{phot}}-z_{\mathrm{spec}}$)/$(1+z_{\mathrm{spec}}$) of $-0.002$ and 0.072 and the fraction of the catastrophic failure ($\delta z >0.5$) of 4.2\% for galaxies with $z_{\mathrm{spec}}$ = 0.8--1.2.
Note that the photometric redshifts are used only for galaxies with no spectroscopic identifications in order to maximize the reliability of the redshift catalog.
The stellar masses were calculated using the best-fit stellar mass-to-luminosity ($M_{\mathrm{s}}$/L) ratio in the observed $K_{\mathrm{s}}$-band.
Typical uncertainties of the stellar masses estimated from the probability distributions in the SED fitting (Figure 2 in K09) show that the stellar mass errors at $z<1.5$ are less than 0.1 and 0.05 dex for stellar masses $M_{\mathrm{s}}$ $>10^{10}$ $M_{\solar}$ and $M_{\mathrm{s}}$ $>10^{11}$ $M_{\solar}$, respectively. 
Those errors include the uncertainty of the photometric redshift for galaxies with no spectroscopic redshift.
It should be mentioned that systematic and random uncertainties (typically 0.2 dex) associated with stellar population models and IMFs are not included in these values.

Figure \ref{fig:MODSsample}(a) shows the stellar masses of galaxies in the MODS deep region at $z=$ 0.6--1.4.
The MODS stellar mass catalog is complete down to at least $M_{\mathrm{s}}$ $\sim10^{9}$ $M_{\solar}$ at these redshifts (shown with the solid line in the figure).
In this paper, we focus on galaxies at $0.8\leq z \leq 1.2$, which are plotted with black dots in the figure.
For those $z\sim1$ galaxies, we plot the distributions of the $K_{\mathrm{s}}$ magnitudes and the rest-frame $U-V$ colors as a function of stellar mass in Figure \ref{fig:MODSsample}(b) and (c), respectively.
There is a tight correlation between $K_{\mathrm{s}}$ magnitudes and stellar masses.
Later, we will limit the sample with $M_{\mathrm{s}}$ $\gtrsim10^{10}$ $M_{\solar}$ considering the $K_{\mathrm{s}}$-band morphological analysis limit of $K_{\mathrm{s}}$(AB) $\sim22.5$ mag (see Section \ref{sect:morph} for the detail). 
The galaxies in the stellar mass range are located in redder part of the bimodal distribution of the rest-frame $U-V$ color in this redshift range.
At $z=$ 0.8--1.2, $\sim$ 80\% (60\%) of the galaxies in the MODS with $M_{\mathrm{s}}$ $\geq$ 3 $\times\ 10^{10}$ $M_{\solar}$ ($M_{\mathrm{s}}$ = 1--3 $\times\ 10^{10}$ $M_{\solar}$) and with no X-ray-detection (see Section \ref{sect:chandra} for identification of X-ray sources) have spectroscopic redshifts.

%%%%%%%%%%%%%%%%%%%%%%%%%%%%%%%%%%%%%%%%%%%%%%%%%%%%%%%%%%%%%%%%%%%%%%%%%%%%%%%
\subsection{MIPS 24$\mu$m flux and IR luminosity of the $z\sim1$ Galaxies}
\label{sect:mips}

In order to estimate IR luminosities, we looked for a \textit{Spitzer}/MIPS (Multiband Imaging Photometer for \textit{Spitzer}) 24 $\mu$m source counterpart for each $z\sim1$ galaxy.
The MIPS 24 $\mu$m source catalog provided by the GOODS team (M. Dickinson et al. in preparation; R. Chary et al. in preparation) contains sources with a 24 $\mu$m flux density ($f_{24}$) larger than 80 $\mu$Jy with 80\% completeness.
The catalog provides not only coordinates of the 24 $\mu$m sources themselves but also those of the IRAC counterparts if exists.
We used the coordinates of the IRAC counterparts if available for the search.
We identified 48 24 $\mu$m counterparts in total for 808 galaxies with $K_{\mathrm{s}}$(Vega) $<24$ mag and $0.8\leq z \leq 1.2$.
In Figure \ref{fig:MODSsample}, the 24 $\mu$m-detected sources are marked with large \textit{filled} circles (red).
Most of them have $K_{\mathrm{s}}$(AB) $<22$ mag, $M_{\mathrm{s}}$ $>10^{10}$ $M_{\solar}$, and bluer colors in the rest-frame $U-V$ than the 24 $\mu$m-undetected galaxies in the same stellar mass range.

Among the $K_{\mathrm{s}}$-selected galaxies with no 24 $\mu$m detection, we examine 112 galaxies with $M_{\mathrm{s}}$ $\geq 1 \times 10^{10}$ $M_{\solar}$ (in our final sample defined in Section \ref{sect:result_nK_ReK}) on the MIPS 24 $\mu$m image (version 0.3) visually in order to estimate how many galaxies would be missed in the 24 $\mu$m catalog by confusion effect. 
We find that 12 out of 112 galaxies are located near bright 24 $\mu$m sources, and two of them have $M_{\mathrm{s}}$ $\geq 3 \times10^{10}$ $M_{\solar}$.
If those 12 galaxies have a 24 $\mu$m flux larger than 80 $\mu$Jy, the fraction of 24 $\mu$m-detected galaxies in our final sample will increase about 3\% (13\%) at $M_{\mathrm{s}}$ $\geq3\times10^{10}$ (= 1--3$\times10^{10}$) $M_{\solar}$.

For the 24 $\mu$m-detected galaxies, we estimate their bolometric IR luminosities $L_{\mathrm{IR}}$ (defined as $L_{\mathrm{IR}}$ = $L$[8--1000 $\mu$m]) from the observed 24 $\mu$m fluxes.
Local IR-luminous ($L_{\mathrm{IR}}$ $>10^{10}$ $L_{\solar}$) galaxies show a tight correlation between 12 $\mu$m flux and $L_{\mathrm{IR}}$ estimated from 12, 25, 60 and 100 $\mu$m MIR flux densities with a scatter of $\sim0.15$ dex (\cite{chary01}).
Following their prescription, we convert the 24 $\mu$m fluxes (in $\mu$Jy) of our $z\sim1$ galaxies to the rest-frame 12 $\mu$m luminosities, and then compute their bolometric IR luminosities in solar luminosity as follows:
\begin{eqnarray}
 \label{eq:nuSnu}
 \nu_{e}S_{\nu_{e}}(12\ {\mu}\mathrm{m})\ &=&\ \nu_{o}\textit{f}_{24}\times4 {\pi}\textit{D}_{L}(z)^{2}, \\
 \label{eq:LIR}
 L_{\mathrm{IR}}\ &=&\ 0.89^{+0.38}_{-0.27}\times[{\nu_{e}}S_{\nu_{e}}(12\ {\mu}\mathrm{m})]^{1.094},
\end{eqnarray}
where ${D}_{L}$ is the luminosity distance corresponding to the redshift $z$, $\nu_{o}$ is the observed frequency corresponding to the wavelength $\lambda=$ 24 $\mu$m, and $\nu_{e}$ is the rest-frame frequency corresponding to the redshift $z$. The symbol $S_{\nu}$ is the monochromatic flux at a frequency $\nu$ expressed in erg s$^{-1}$ Hz$^{-1}$. 
We need to note that the IR luminosity estimated using the rest-frame 12 $\mu$m flux density has an uncertainty of $\sim$ 0.3 dex due to the uncertainty of the shapes of the IR SED (see \cite{lefloch05}; \cite{marc06}).

The 24 $\mu$m flux density limit of $f_{24}$ $=80$ $\mu$Jy corresponds to the IR luminosity $L_{\mathrm{IR}}$ of $0.6\times 10^{11}$, $1.1\times10^{11}$ and $2.1\times 10^{11}$ $L_{\solar}$, and SFR of $\sim10$, 15, and 35 $M_{\solar}$ $\mathrm{yr}^{-1}$ at $z=$ 0.8, 1.0 and 1.2, respectively.
In this paper, we regard all the 24 $\mu$m-detected ($f_{24}$ $\geq$ 80 $\mu$Jy) galaxies as LIRGs while 24 $\mu$m-undetected ($f_{24}$ $<$ 80 $\mu$Jy) galaxies as non-LIRGs.
We need to note that the LIRG identification is incomplete at $z=$ 1.0--1.2 in the strict definition of LIRGs with $L_{\mathrm{IR}}$ $=$ $10^{11}$--$10^{12}$ $L_{\solar}$.

%%%%%%%%%%%%%%%%%%%%%%%%%%%%%%%%%%%%%%%%%%%%%%%%%%%%%%%%%%%%%%%%%%%%%%%%%%%%%%%
\subsection{Identifying Chandra X-ray Sources}
\label{sect:chandra}

A part of the $z\sim 1$ galaxies is also detected in X-ray; 18 out of 808 galaxies with $K_{\mathrm{s}}$ $< 24$ mag and $0.8 \leq z \leq 1.2$ are detected in X-ray source catalog from the \textit{Chandra} X-ray Observatory deep survey image (\cite{alex03}).
In Figure \ref{fig:MODSsample}, those X-ray-detected sources are marked with \textit{open} circles. We checked those sources in the ACS $z_{850}$-band image, and all but one source are extended.

The evaluation of morphologies of those galaxies may be affected by bright point source of type I AGN.
Also, if the SED of a galaxy is strongly contaminated by an AGN, no reliable stellar mass could be obtained without considering the effect of AGNs in the SED fitting. 
Moreover, MIR fluxes could not be a good indicator of SFR if their MIR radiations are mainly powered by UV radiation from AGNs.
We will indicate those X-ray-detected galaxies with different symbols to the others in later figures, and exclude from the statistics.

%%%%%%%%%%%%%%%%%%%%%%%%%%%%%%%%%%%%%%%%%%%%%%%%%%%%%%%%%%%%%%%%%%%%%%%%%%%%%%%
\section{Morphological Analysis}
\label{sect:morph}

%%%%%%%%%%%%%%%%%%%%%%%%%%%%%%%%%%%%%%%%%%%%%%%%%%%%%%%%%%%%%%%%%%%%%%%%%%%%%%%
\subsection{$\rm{S\acute{e}rsic}$ profile fitting: Method}
\label{sect:galfit_method}

We evaluate the rest-frame NIR morphologies of the $z\sim1$ galaxies using the MODS $K_{\mathrm{s}}$-band image which is constructed from good seeing images.
The MODS deep $K_{\mathrm{s}}$-band data consists of images with various seeing sizes, FWHMs of from \timeform{0''.4} to \timeform{1''.2}.
The image used in K09 was the deepest one created by combining images with the FWHM smaller than \timeform{0''.8}, which has a total integration time of $\sim28$ hours, 3$\sigma$ limiting magnitude of $K_{\mathrm{s}}$(AB) $\sim26.4$ mag for point sources, and FWHM of the PSF of $\sim$ \timeform{0''.46}.
Here, for the morphological analysis, we use a \textit{shallower} but \textit{sharper} image constructed from images with the FWHM $<$ \timeform{0''.5}, which gives a total integration time of $\sim$ 16 hours and 3$\sigma$ limiting magnitude of $K_{\mathrm{s}}$(AB) $\sim 26.0$ mag for a point source, and the final FWHM of \timeform{0''.40}, which corresponds to a physical scale of $\sim$ 3.2 kpc at $z=1$ in the cosmology we use in this paper.

In order to describe morphologies of the $z\sim1$ galaxies, we use a single component $\rm{S\acute{e}rsic}$ model (\cite{sersic68}; see also \cite{graham05} for detail of the model).
It describes the radial surface brightness profile of a galaxy by a function given by,

\begin{eqnarray}
 \label{eq:sersic}
 I(r) &=& I(r_{\mathrm{e}}) \exp \left\{-b_{n} \left[ \left( \frac{r}{r_{\mathrm{e}}} \right)^{1/n}-1 \right] \right\}, \\
 r &=& \left[ (x-x_{\mathrm{c}})^{2} + \left(\frac{y-y_{\mathrm{c}}}{q}\right)^{2} \right] ^{1/2}, %{\frac{1}{2}},
\end{eqnarray}
where $x$ and $y$ are aligned with the semimajor and semiminor axes, $r$ is the elliptical radial distance at a pixel ($x, y$) from the center of a source ($x_{\mathrm{c}},y_{\mathrm{c}}$), $q$ is the axial ratio of the semiminor to the semimajor axis radius, $r_{\mathrm{e}}$ is the major axis half-light radius or effective radius which contains the half of the total flux, and $n$ is the $\rm{S\acute{e}rsic}$ index which determines how the light profile concentrates around the center and how the profile extends to the outskirt.
The $b_{n}$ is a normalization constant which is a function of $n$ and is chosen so that $r_{\mathrm{e}}$ is equivalent to the half-light radius.
The $\rm{S\acute{e}rsic}$ model includes a wide range of profiles of local galaxies such as the exponential ($n$ $=1$) and de Vaucouleurs ($n$ $=4$) models.
In fact, a tight correlation between $n$ and Hubble T-type ($T$) is observed in the local universe (\cite{ravin04}) and even at $z\lesssim1$ (\cite{pann06}); on average, galaxies with $n$ $<$ 2--2.5 are mostly disk-like objects ($T\gtrsim2$) while galaxies with $n$ $>$ 2--2.5 are mostly spheroids (elliptical and bulge-dominated galaxies; $T\lesssim2$).

We use the two-dimensional surface-brightness profile fitting code, \textsc{Galfit} ver.2.0.3c (\cite{peng02}).
\textsc{Galfit} convolves a two-dimensional model profile with a point-spread function (PSF) defined by user, and minimizes $\chi^{2}$ residuals between the model profile and a real galaxy profile with a Levenberg-Marquardt algorithm. 
In the $\chi^{2}$ minimization, there are 7 free parameters: $\rm{S\acute{e}rsic}$ parameters ($n$ and $r_{\mathrm{e}}$), total magnitude, semiminor-to-semimajor axial ratio $q$, position angle $PA$, and the center position ($x_{\mathrm{c}},y_{\mathrm{c}}$) of a galaxy.
Initial guesses for those parameters except for $n$ (being set to 1.5) were estimated based on the output parameters from \textsc{SExtractor} (\cite{bertin96}) object detection code.
To estimate $\chi^{2}$ correctly, we take into account the pixel-to-pixel signal-to-noise ratio. 
The dominant source of noise for NIR imaging data is Poissonian noise from sky background radiation.
Therefore we use the square root of the exposure map as pixel-to-pixel weights in the $\chi^{2}$ minimization.

The PSF used in this paper is obtained by stacking about 10 spectroscopically-identified isolated stars.
The stars are distributed over the $K_{\mathrm{s}}$-band image, and their FWHMs are quite uniform (3.42 $\pm$ 0.02 pixel).
We therefore ignored the dependence on the position in the image, and combined all the stars to create a composite PSF, which is applied to all the $z\sim1$ galaxies.

For a fitting region, one must have a sufficiently wide image size of a galaxy to cover the outskirts of the profile in the calculation.
Following the galaxy two-dimensional profile fitting procedure used in the Galaxy Evolution from Morphology and SEDs (GEMS) survey (\cite{haussler07}), we use an image stamp for each galaxy with a size (height and width) of 2.5 times the semimajor axial length of the Kron elliptical aperture derived from the \textsc{SExtractor} output which is expected to contain more than 90\% of the total flux of the galaxy regardless of brightness (\cite{kron80}).
Neighboring galaxies around the galaxy of interest were masked and excluded from the fitting, or were fitted simultaneously if they are so close to the galaxy of interest. 
We determined whether a neighbor should be masked out or fitted simultaneously based on their Kron elliptical apertures same as in \citet{haussler07}.
First, the apertures are enlarged by increasing the semiminor and the semimajor axis radii by a factor of 1.5, and then if the neighbor's extended aperture overlaps the extended aperture of the galaxy of interest, the neighbor is fitted simultaneously; otherwise, the neighbor is masked out with the extended aperture.
Neighbors with $K_{\mathrm{s}}$(AB) $>22.5$ mag are masked out regardless of the overlap because the profile fitting for them would not be reliable (see the next subsection).
Note that a difference in the treatment of neighbors (masked or fitted) has little effect on the fitting result for a galaxy of interest, except for galaxies with very bright neighbors, which suffer flux contamination from the neighbors and their fittings fail or derive obviously unreliable results with bright residuals.

%%%%%%%%%%%%%%%%%%%%%%%%%%%%%%%%%%%%%%%%%%%%%%%%%%%%%%%%%%%%%%%%%%%%%%%%%%%%%%%
\subsection{$\rm{S\acute{e}rsic}$ profile fitting: Simulation}
\label{sect:galfit_sim}

\begin{figure}
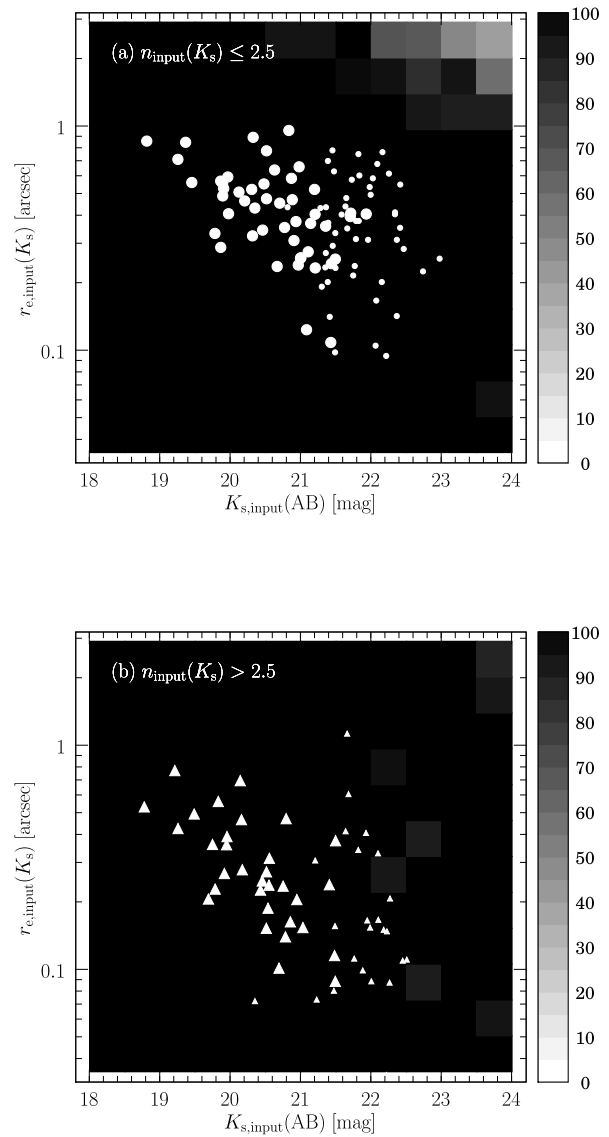

  \begin{center}
    \FigureFile(80mm,80mm){figure02a.eps}
    \FigureFile(80mm,80mm){figure02b.eps}
  \end{center}
  \caption{Detection completeness map for galaxies in the MODS $K_{\mathrm{s}}$-band image estimated from the artificial galaxy simulation. (a) The map for artificial galaxies with $n_{\mathrm{input}}(K_{\mathrm{s}})$ $\leq2.5$. The grey-scale map indicates the simulation result. Overplotted large (small) \textit{circles} show the observed effective radii and $K_{\mathrm{s}}$ magnitudes of the $z\sim1$ galaxies with $M_{\mathrm{s}}$ $\geq 3\times10^{10}$ $M_{\solar}$ ($M_{\mathrm{s}}$ = 1--3 $\times10^{10}$ $M_{\solar}$) and $n(K_{\mathrm{s}})$ $\leq2.5$ in the MODS. (b) Similar to the panel (a), but for galaxies with $n(K_{\mathrm{s}})$ $>2.5$.}
  \label{fig:det_comp_disk}
\end{figure}

\begin{figure}
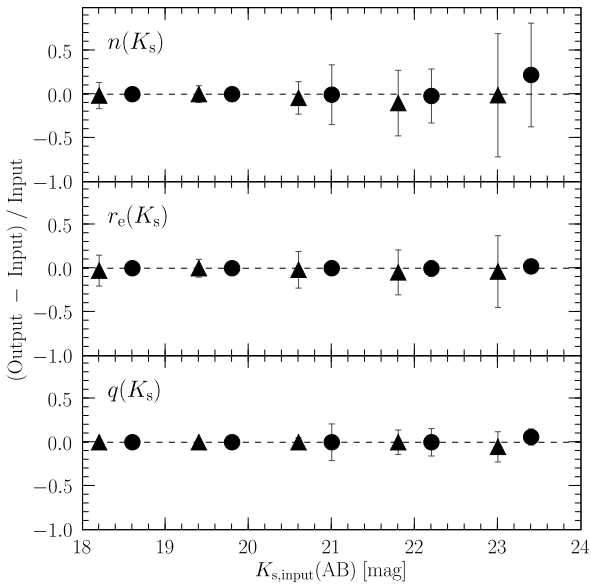

  \begin{center}
    \FigureFile(80mm,80mm){figure03.eps}
  \end{center}
  \caption{Magnitude dependences of the difference between input and output structural parameters for artificial galaxies (\textit{circles} for those with $n_{\mathrm{input}}[K_{\mathrm{s}}]$ $\leq2.5$ and \textit{triangles} for those with $n_{\mathrm{input}}[K_{\mathrm{s}}]$ $>2.5$). From the \textit{top} to \textit{bottom} panel, the median difference and $1\sigma$ scatter are shown for $\rm{S\acute{e}rsic}$ indices $n(K_{\mathrm{s}})$, effective radii $r_{\mathrm{e}}(K_{\mathrm{s}})$, and axial ratios $q(K_{\mathrm{s}})$. Among the galaxies detected, we used only those with the similar $K_{\mathrm{s}}$ magnitude ($\pm0.2$ mag) and the effective radius (80--120\%) compared to the observed $z\sim1$ galaxies.}
  \label{fig:morph_mag_depend}
\end{figure}

\begin{figure}
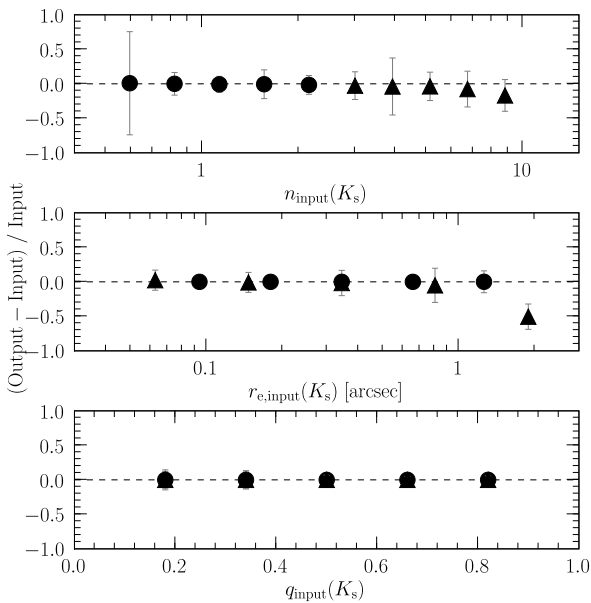

  \begin{center}
    \FigureFile(80mm,80mm){figure04.eps}
  \end{center}
  \caption{Difference between input and output values for each structural parameter as a function of input value for artificial galaxies. From the \textit{top} to \textit{bottom} panel, the median difference and $1\sigma$ scatter are shown for $\rm{S\acute{e}rsic}$ indices $n(K_{\mathrm{s}})$, effective radii $r_{\mathrm{e}}(K_{\mathrm{s}})$, and axial ratios $q(K_{\mathrm{s}})$. The symbols are the same as in Figure \ref{fig:morph_mag_depend}. Among the galaxies used in Figure \ref{fig:morph_mag_depend}, we used only those with $K_{\mathrm{s,input}}\rm(AB)\leq22.5$ mag.}
  \label{fig:input_residual}
\end{figure}

To quantify how reliably we evaluate morphology of galaxies with the MODS $K_{\mathrm{s}}$-band image, we perform a Monte Carlo simulation with $\sim$ 1000 artificial galaxies.
\textsc{Galfit} is used to generate artificial galaxy images with parameters given randomly in the following ranges:
18 $\leqq$ $K_{\mathrm{s}}$(AB) $\leqq$ 24 mag, 0.3 $\leqq$ $r_{\mathrm{e}}$ $\leqq$ 25 pixel (equivalent to \timeform{0.''035}--\timeform{2.''9} or 0.3--23 kpc at $z=1$), 0.5 $\leqq$ $n(K_{\mathrm{s}})$ $\leqq$ 10, 0.1 $\leqq$ $q$ $\leqq$ 0.9, and 0 $\leqq$ $PA$ $\leqq$ 180 deg.
Each artificial galaxy is convolved with the PSF used for the real galaxies, and is embedded randomly in the $K_{\mathrm{s}}$-band image.
\textsc{SExtractor} is used to detect them and measure apparent shapes, which are used as initial guesses for \textsc{Galfit} in the same manner as for the real galaxies.

Using the number of galaxies detected among galaxies embedded, we first show detection completeness as a function of input magnitude and effective radius for low-$n_{\mathrm{input}}(K_{\mathrm{s}})$ ($n_{\mathrm{input}}$[$K_{\mathrm{s}}$]$\leq2.5$) sample and high-$n_{\mathrm{input}}(K_{\mathrm{s}})$ ($n_{\mathrm{input}}[K_{\mathrm{s}}]$ $>2.5$) sample in Figure \ref{fig:det_comp_disk}(a) and (b), respectively.
Also shown in both panels is the distribution of the $K_{\mathrm{s}}$ magnitude (MAG\_AUTO from \textsc{SExtractor}) and effective radius evaluated with \textsc{Galfit} of the $z\sim1$ galaxies with $M_{\mathrm{s}}$ $\geq1\times10^{10}$ $M_{\solar}$ according to $n(K_{\mathrm{s}})$.
Although faint and large galaxies (located at top-right in Figure \ref{fig:det_comp_disk}) can be missed at $K_{\mathrm{s}}$(AB) $\geq22.5$ mag in the low-$n(K_{\mathrm{s}})$ sample, all the detected $z\sim1$ galaxies are located well below the detection limit.

Comparing input parameters with measured ones for the detected artificial galaxies, we estimate the uncertainty of the profile fitting parameters.
For this analysis, we only select the artificial galaxies having measured $K_{\mathrm{s}}$-band magnitudes and effective radii within 20\% of those of the observed $z\sim1$ galaxies (shown in Figure \ref{fig:det_comp_disk} with filled symbols).
Figure \ref{fig:morph_mag_depend} shows the differences between the input and measured parameters as a function of $K_{\mathrm{s}}$ input magnitude.
There is no systematic offset in any parameters up to $K_{\mathrm{s,input}}$(AB) $=22.5$ mag.
At $K_{\mathrm{s,input}}$(AB) $\sim22.5$ mag, we find that low-$n_{\mathrm{input}}(K_{\mathrm{s}})$ galaxies tend to have $-1(\pm34)$\% lower $\rm{S\acute{e}rsic}$ index, $0(\pm9)$\% lower size, and $1(\pm14)$\% larger axial ratio, and high-$n_{\mathrm{input}}(K_{\mathrm{s}})$ galaxies tend to have $-3(\pm45)$\% lower $\rm{S\acute{e}rsic}$ index, $-4(\pm29)$\% lower size, and $-5(\pm16)$\% lower axial ratio.

Finally, we examine differences between input and output values for each parameter as a function of input value of the galaxies with $K_{\mathrm{s}}$(AB) $\leq22.5$ mag, which are shown in Figure \ref{fig:input_residual}.
We find no significant trend in each parameter for the low-$n_{\mathrm{input}}(K_{\mathrm{s}})$ sample.
On the contrary, for the high-$n_{\mathrm{input}}(K_{\mathrm{s}})$ sample we find a systematic offset of $\Delta n/n$ $\sim-15$\% and $\Delta r_{\mathrm{e}}/r_{\mathrm{e}}$ $\sim-10$\% for galaxies with $n_{\mathrm{input}}(K_{\mathrm{s}})$ $\gtrsim$ 8.
Although those offsets less affect to our morphological separation criterion of $n(K_{\mathrm{s}})$ $=2.5$ (described in Section \ref{sect:result_nK_ReK}), we have to keep in mind the offset in size when we study the sizes of the high-$n(K_{\mathrm{s}})$ galaxies.

Based on the results of the simulation with artificial galaxies, we find that the morphological analysis using \textsc{Galfit} with the MODS deep $K_{\mathrm{s}}$-band image is reliable up to $K_{\mathrm{s}}$(AB) $\sim22.5$ mag.
Therefore, in later sections we will limit our $z\sim1$ morphological sample at the stellar mass of $M_{\mathrm{s}}$ $\sim10^{10}$ $M_{\solar}$ corresponding to that magnitude.

%%%%%%%%%%%%%%%%%%%%%%%%%%%%%%%%%%%%%%%%%%%%%%%%%%%%%%%%%%%%%%%%%%%%%%%%%%%%%%%
\subsection{Non-parametric morphological analysis}
\label{sect:CA}

We perform an another morphological analysis based on a non-parametric method: Concentration-Asymmetry ($C$-$A$) analysis (\cite{abraham94}; \cite{abraham96}; \cite{conselice00}; \cite{conselice03}).
Unlike \textsc{Galfit}, the $C$-$A$ analysis does not require information on the shape of the PSF and assumptions on the shape of the light profile.
Therefore, we use this method in parallel with the $\rm{S\acute{e}rsic}$ profile fitting.

The concentration index ($C$) is an indicator how the light profile of a galaxy concentrates around the center.
The definition used in this paper is,
\begin{equation}
 C = 5 \textrm{log}_{10} \left( \frac{r_{80}}{r_{20}} \right),
\end{equation}
where $r_{80}$ and $r_{20}$ are the radii which contain 80\% and 20\% of the total flux, respectively. The total flux is defined as MAG\_AUTO derived from the \textsc{SExtractor} output.
This parameter is related to $n$ of $\rm{S\acute{e}rsic}$ profile and there is certainly a correlation between these two parameters.

The asymmetry index ($A$) is an indicator how the two-dimensional light profile is disturbed, which is defined as,
\begin{equation}
\label{eq:asym}
 A = \frac{1}{2} \left[ \textrm{min} \left( \frac{\Sigma|I_{0}-I_{180}|}{\Sigma|I_{0}|} \right) - \textrm{min} \left( \frac{\Sigma|B_{0}-B_{180}|}{\Sigma|I_{0}|} \right) \right],
\end{equation}
where $I_{0}$ and $I_{180}$ are the intensity of each pixel of the image and of the image rotated by 180$^{\circ}$ around the galaxy centroid, respectively.
The symbols $B_{0}$ and $B_{180}$ have the similar definitions but for the background region. 
The $A$ is defined as the minimum value with varying the position of the center of rotation around the original galaxy centroid.
The summation taken in the area is defined using the Kron elliptical aperture used in Section \ref{sect:galfit_method}.
We found that the contribution of sky background, the term $\Sigma|B_{0}-B_{180}|$, correlates well to the elliptical aperture size, and is less dependent on the position on the image.
Therefore, we assume a universal background over the image and scale it according to the aperture size of galaxies.

%%%%%%%%%%%%%%%%%%%%%%%%%%%%%%%%%%%%%%%%%%%%%%%%%%%%%%%%%%%%%%%%%%%%%%%%%%%%%%%
\section{Results}
\label{sect:result}

%%%%%%%%%%%%%%%%%%%%%%%%%%%%%%%%%%%%%%%%%%%%%%%%%%%%%%%%%%%%%%%%%%%%%%%%%%%%%%%
\subsection{Rest-frame NIR morphologies of z $\sim$ 1 LIRGs}
\label{sect:result_nK_ReK}

\begin{table*}
  \begin{center}
  \caption{The number of sample galaxies in this paper\footnotemark[$*$]}
  \label{tab:sample}
    \begin{tabular}{l|cccc|cccc}
\hline \hline
 & \multicolumn{4}{c|}{$M_{\mathrm{s}}$ = 1--3 $\times10^{10}$$M_{\solar}$} & \multicolumn{4}{c}{$M_{\mathrm{s}}$ $\geq 3 \times10^{10}$$M_{\solar}$} \\
 & \multicolumn{2}{c}{No X-ray} & \multicolumn{2}{c|}{All} & \multicolumn{2}{c}{No X-ray} & \multicolumn{2}{c}{All} \\
\hline
LIRGs with $n$[$K_{\mathrm{s}}$]$\leq2.5$	 & 7  & (6)  & 7  & (6)  & 27  & (21)  & 30  & (24)  \\
LIRGs with $n$[$K_{\mathrm{s}}$]$>2.5$	 & 2  & (1)  & 3  & (2)  & 1  & (1)  & 3  & (3)  \\
Non-LIRGs with $n$[$K_{\mathrm{s}}$]$\leq2.5$	 & 43  & (29)  & 43  & (29)  & 15  & (8)  & 18  & (11)  \\
Non-LIRGs with $n$[$K_{\mathrm{s}}$]$>2.5$	 & 20  & (6)  & 22  & (8)  & 24  & (21)  & 29  & (26)  \\
\hline
Total	 & 72  & (42)  & 75  & (45)  & 67  & (51)  & 80  & (64)  \\
\hline
\multicolumn{4}{@{}l@{}}{\hbox to 0pt{\parbox{105mm}{\footnotesize
 \footnotemark[$*$]In the parentheses, numbers of galaxies with spectroscopic redshift are given.
}\hss}}

    \end{tabular}
  \end{center}
\end{table*}

\begin{figure}
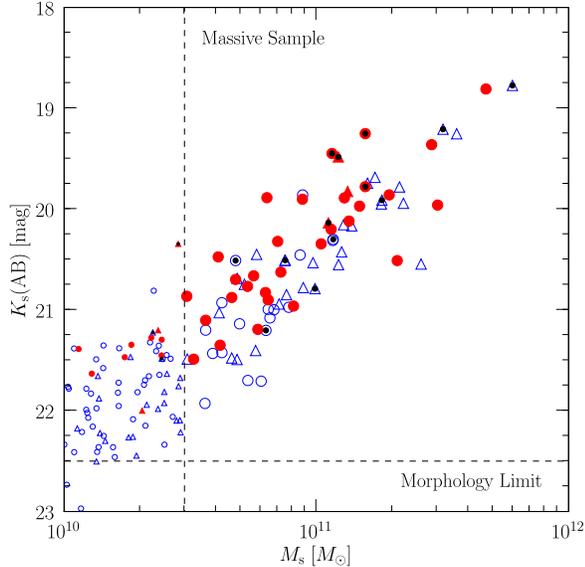

  \begin{center}
    \FigureFile(80mm,80mm){figure05.eps}
  \end{center}
  \caption{{$K_{\mathrm{s}}$} magnitude distribution of $z\sim1$ galaxies as a function of stellar mass. The \textit{filled} symbols show the LIRGs while \textit{open} symbols show the non-LIRGs. The sample is divided by the $\rm{S\acute{e}rsic}$ index: \textit{circles} for low-$n(K_{\mathrm{s}})$ ($n$[$K_{\mathrm{s}}$] $\leq2.5$) galaxies and \textit{triangles} for high-$n(K_{\mathrm{s}})$ ($n$[$K_{\mathrm{s}}$] $>2.5$) galaxies. The X-ray-detected sample is marked with black dots. The horizontal \textit{dashed} line represents $K_{\mathrm{s}}$(AB) = 22.5 mag, which indicates the limit of morphological analysis. This magnitude corresponds to the stellar mass of $\sim1\times10^{10}$ $M_{\solar}$ which is adopted as our stellar mass-limited morphological sample criterion. The vertical \textit{dashed} line separates less massive ($M_{\mathrm{s}}$ = 1--3 $\times 10^{10}$ $M_{\solar}$) and massive ($M_{\mathrm{s}}$ $\geq 3 \times 10^{10}$ $M_{\solar}$) galaxies. The less massive galaxies are indicated with smaller symbols.}
  \label{fig:Ms_Ks_morph}
\end{figure}

\begin{figure}
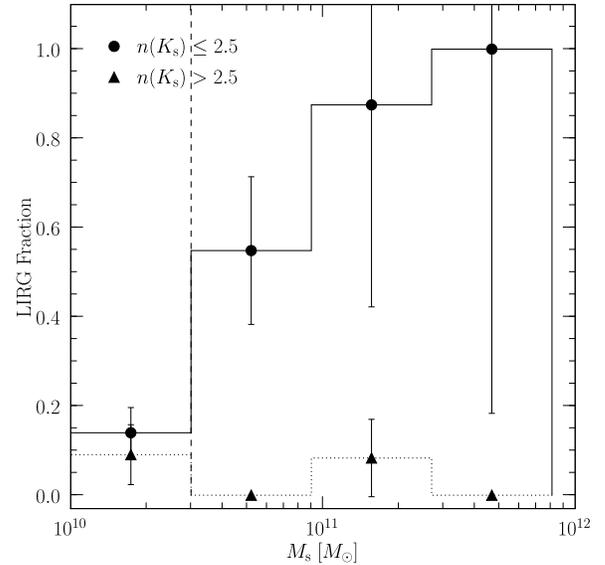

  \begin{center}
    \FigureFile(80mm,80mm){figure06.eps}
  \end{center}
  \caption{Fraction of LIRGs (galaxies with $f_{24}$ $\geq$ 80 $\mu$Jy) as a function of stellar mass. The samples are split into two groups ($n$[$K_{\mathrm{s}}$] $\leq2.5$: \textit{circles} and $n$[$K_{\mathrm{s}}$] $>2.5$: \textit{triangles}). All the X-ray-detected galaxies are excluded. The error bars account for Poissonian errors only.}
  \label{fig:Ms_LIRG_fraction}
\end{figure}

The morphological simulation shows that the MODS $K_{\mathrm{s}}$-band image allows us to evaluate a light profile of galaxies with $K_{\mathrm{s}}$(AB) $\sim22.5$ mag or brighter (Section \ref{sect:galfit_sim}). 
As seen in Figure \ref{fig:MODSsample}(b), this $K_{\mathrm{s}}$ magnitude limit corresponds to $M_{\mathrm{s}}$ $\sim 10^{10}$ $M_{\solar}$ at $z\sim1$.
We therefore apply a stellar mass cut of $M_{\mathrm{s}}$ $= 1\times10^{10}$ $M_{\solar}$ to the $z\sim1$ galaxy sample to define a morphological sample, which contains 139 galaxies in total.
Considering the scatter in $K_{\mathrm{s}}$ magnitude at $M_{\mathrm{s}}$ $= 1\times 10^{10}$ $M_{\solar}$ (representing a variety of $M_{\mathrm{s}}$/L), we also construct a more conservative morphological sample containing galaxies with $M_{\mathrm{s}}$ $\geq3\times10^{10}$ $M_{\solar}$, to which 67 out of 139 galaxies belongs.
We refer the galaxies with $M_{\mathrm{s}}$ $\geq3\times10^{10}$ $M_{\solar}$ ($M_{\mathrm{s}}$ = 1--3 $\times10^{10}$ $M_{\solar}$) as massive (less massive) galaxies.
Our morphological sample achieves high completeness thanks to the very deep imaging data; 100\% and 96\% for the massive and less massive sample, respectively.
We summarize in Table \ref{tab:sample} the numbers of the $z\sim1$ galaxy morphological sub-sample in the MODS deep region for stellar mass cuts of $M_{\mathrm{s}}$ = 1--3 $\times10^{10}$ $M_{\solar}$ and $M_{\mathrm{s}}$ $\geq3\times10^{10}$ $M_{\solar}$.
In the parentheses, the numbers of galaxies with spectroscopic redshift are shown.
The spectroscopic redshifts are available for about 60\% and 80\% of less massive and massive galaxies, respectively.
The uncertainty of the photometric redshift estimates of 0.072 (Section \ref{sect:mods}) brings about 1--2\% uncertainties in physical size ($R_{\mathrm{e}}$) estimates, which is sufficiently small for our study.
In Appendices we show the properties and images used in this study for all the $z\sim1$ sample.

Figure \ref{fig:Ms_Ks_morph} summarizes the morphological evaluation for the $z\sim1$ galaxies in the stellar mass--$K_{\mathrm{s}}$ magnitude diagram.
The sample is divided by their IR luminosities (LIRGs and non-LIRGs) and $\rm{S\acute{e}rsic}$ indices (less concentrated system: low-$n[K_{\mathrm{s}}]$ and highly concentrated system: high-$n[K_{\mathrm{s}}]$) in the figure.
The X-ray-detected sample is marked with black dots.
For $M_{\mathrm{s}}$ $< 3\times10^{10}$ $M_{\solar}$, the sample is shown with small symbols.
In addition to the fact that the LIRGs are popular at $M_{\mathrm{s}}$ $>10^{10}$ $M_{\solar}$ and bright in the $K_{\mathrm{s}}$-band as already seen in Figure \ref{fig:MODSsample}(b), most of them are classified in $n(K_{\mathrm{s}})$ $\leq2.5$.
In Figure \ref{fig:Ms_LIRG_fraction}, these trends are shown quantitatively.
Although with small sample size of the LIRGs at the most massive bin, the fraction of LIRGs in the sample increases with stellar mass, and most of them have $n(K_{\mathrm{s}})$ $\leq2.5$ while the LIRGs with $n(K_{\mathrm{s}})$ $>2.5$ are rare.

\begin{figure*}
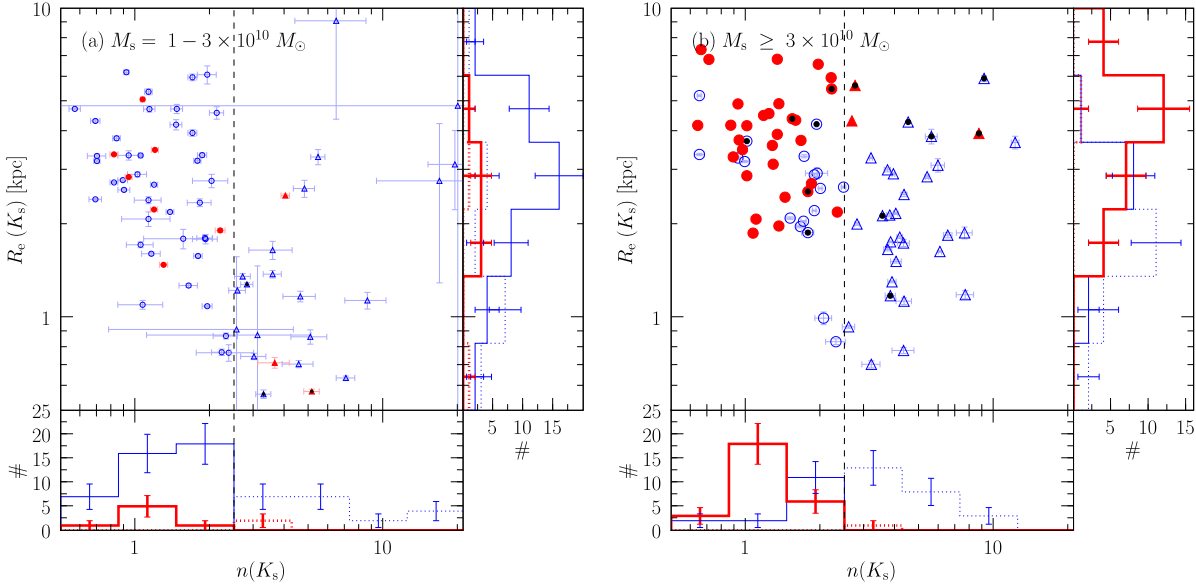

  \begin{center}
    \begin{tabular}{c}
      \FigureFile(80mm,80mm){figure07a.eps}
      \FigureFile(80mm,80mm){figure07b.eps}
    \end{tabular}
  \end{center}
  \caption{Structural parameters ($\rm{S\acute{e}rsic}$ indices and effective radii in physical scale along the semimajor axis) measured in the MODS deep $K_{\mathrm{s}}$-band image for the $z\sim1$ galaxies with (a) $M_{\mathrm{s}}$ = 1--3 $\times10^{10}$ $M_{\solar}$, and with (b) $M_{\mathrm{s}}$ $\geq3\times10^{10}$ $M_{\solar}$. The symbols are the same as in Figure \ref{fig:Ms_Ks_morph}. The error bars indicate the uncertainties of the fitting. The vertical dashed line separates less concentrated and highly concentrated systems ($n$[$K_{\mathrm{s}}$] $\leq2.5$ is considered as a less concentrated system). In each figure, the bottom and right panels show numbers of the samples at a given stellar mass range as a function of $n(K_{\mathrm{s}})$ and $R_{\mathrm{e}}$($K_{\mathrm{s}}$), respectively. Thick (thin) line corresponds to the numbers of the LIRGs (non-LIRGs), and solid (dotted) line corresponds to the numbers of the low-$n(K_{\mathrm{s}})$ (high-$n[K_{\mathrm{s}}]$) sample. The error bars of the histograms account for Poissonian errors only. All the X-ray-detected galaxies are excluded from the counts.}
  \label{fig:nK_ReK}
\end{figure*}

Figure \ref{fig:nK_ReK}(a) and (b) show the distribution of the structural parameters, the $\rm{S\acute{e}rsic}$ indices $n(K_{\mathrm{s}})$ and the physical sizes (effective radii in kpc) along the semimajor axis $R_{\mathrm{e}}(K_{\mathrm{s}})$, for galaxies with $M_{\mathrm{s}}$ = 1--3 $\times10^{10}$ $M_{\solar}$ and with $M_{\mathrm{s}}$ $\geq3\times10^{10}$ $M_{\solar}$, respectively.
In each figure, the bottom and right panels show the respective distributions of the parameters, where the X-ray-detected galaxies are excluded from the counts.
Most of the LIRGs are well described with exponential disk-like profiles ($\langle n[K_{\mathrm{s}}$]$\rangle$ $\sim$ 1.2).
With our separation criterion of $n(K_{\mathrm{s}})$ $=2.5$ in the massive ($M_{\mathrm{s}}$ $\geq3\times10^{10}$ $M_{\solar}$) sample, 96\% (27/[27+1]) of the LIRGs are classified in less concentrated (low-$n[K_{\mathrm{s}}]$) system while only one galaxy ($\sim$3\%) is in highly concentrated (high-$n[K_{\mathrm{s}}]$), or bulge-dominated system.
As seen in the bottom panel of Figure \ref{fig:nK_ReK}(b), the LIRGs have a skewed $n(K_{\mathrm{s}})$-distribution where most of the LIRGs are distributed between $n(K_{\mathrm{s}})$ = 1--2, and therefore the fraction is not changed significantly even if we use a different separation criterion; about 89\% and 100\% for the criterion $n(K_{\mathrm{s}})$ $=2.0$ and $3.0$, respectively. 
\citet{mel08} reported the similar result to ours by AO-supported NIR observations that two thirds ($\sim67$\%) of $z\sim1$ LIRGs are disk-like galaxies according to their visual morphological classifications.
Despite various differences between our study and \citet{mel08} such as the spatial resolution of data (seeing-limited or AO-supported) or the morphological classification methods (based on the profile fitting or visual appearance), 
our study confirms quantitatively with larger number of sample that most of the $z\sim1$ LIRGs have disk-like morphologies.

If we examine the fraction among the massive galaxies, we find that about 64\% (27/[27+15]) of the low-$n(K_{\mathrm{s}})$ galaxies are LIRGs among the galaxies with $M_{\mathrm{s}}$ $\geq 3 \times 10^{10}$ $M_{\solar}$. 
As seen in Figure \ref{fig:Ms_LIRG_fraction} as well, the LIRG fraction in the low-$n(K_{\mathrm{s}})$ sample increases with stellar mass.
At $z\sim1$, the disk-like massive galaxies have surprisingly high fraction of LIRGs.
In fact, the residual images after subtracting best fit $\rm{S\acute{e}rsic}$ component show clear spiral structure for some of $z\sim1$ galaxies.

As for the sizes, the right panel of Figure \ref{fig:nK_ReK}(b) shows that the LIRGs have larger $R_{\mathrm{e}}$($K_{\mathrm{s}}$) than the non-LIRGs on average.
Comparing the low-$n(K_{\mathrm{s}})$ galaxies in between LIRGs and non-LIRGs, a mean size and a sample scatter of the LIRGs is $3.8\pm1.6$ kpc while that of the non-LIRGs is $2.6\pm1.0$ kpc.
The high-$n(K_{\mathrm{s}})$ non-LIRGs have a wide $R_{\mathrm{e}}$($K_{\mathrm{s}}$)-distribution and some of them have the sizes around 1 kpc.
In the MODS deep data, a FWHM of the PSF is \timeform{0''.40}, which corresponds to $R_{\mathrm{e}}$($K_{\mathrm{s}}$) $\sim$ 2 kpc at $z=1$. 
Since \textsc{Galfit} takes into account the PSF in the profile fitting, sizes smaller than the PSF size can be evaluated in principle. 
However, the morphological analysis of those unresolved galaxies can be easily affected by what PSF (size and shape) is assumed in the fitting, and their resultant sizes would have larger uncertainties than the other larger galaxies.
Therefore, we have to pay attention to those small ($R_{\mathrm{e}}$[$K_{\mathrm{s}}$] $\leq$ 2 kpc) galaxies when discussing their sizes.
However, for the LIRGs, most of them are much larger than the PSF size and their profiles must be derived reliably.

%%%%%%%%%%%%%%%%%%%%%%%%%%%%%%%%%%%%%%%%%%%%%%%%%%%%%%%%%%%%%%%%%%%%%%%%%%%%%%%
\subsection{Rest-frame NIR Stellar Mass--Size Relation}
\label{sect:Ms_Re_K}

\begin{figure*}
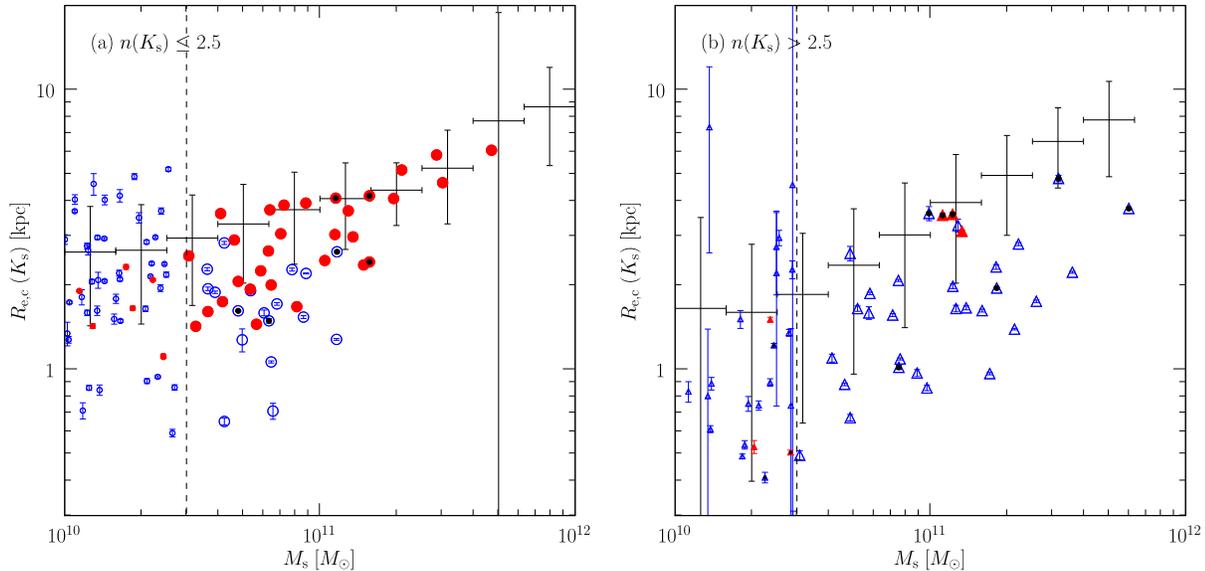

  \begin{center}
    \begin{tabular}{c}
      \FigureFile(80mm,80mm){figure08a.eps}
      \FigureFile(80mm,80mm){figure08b.eps}
    \end{tabular}
  \end{center}
  \caption{The circularized effective radius of the $z\sim1$ galaxies with (a) $n(K_{\mathrm{s}})$ $\leq 2.5$, and with (b) $n(K_{\mathrm{s}})$ $> 2.5$ as a function of stellar mass. The symbols are the same as Figure \ref{fig:Ms_Ks_morph}. The error bars indicate the uncertainties of the fitting. The median and dispersion of the distribution of effective radius in $z$-band of local late-type ($n[z]\leq2.5$) and early-type ($n[z]>2.5$) galaxies (redshifts of $\sim0.062$) derived from VAGC (\cite{blanton05}) are overplotted as a function of stellar mass.}
\label{fig:Ms_RecK}
\end{figure*}

We show the stellar mass--size distribution in the $K_{\mathrm{s}}$-band (rest-frame $J$-band) in Figure \ref{fig:Ms_RecK}(a) and (b) for the low-$n$($K_{\mathrm{s}}$) and high-$n$($K_{\mathrm{s}}$) sample, respectively.
The sizes are shown as circularized effective radii, $R_{\mathrm{e,c}}$ $=$ $R_{\mathrm{e}}$ $\times$ $q^{1/2}$, where $q$ is the intrinsic semiminor-to-semimajor axial ratio of the galaxy.
The circularized method yields large difference in size between the LIRGs and non-LIRGs when compared to those derived from the semimajor axial radius of the galaxies which is shown in Figure \ref{fig:nK_ReK}.
To compare those sizes with those of local galaxies, we construct a local galaxy catalog including spectroscopic redshifts, stellar masses, and $\rm{S\acute{e}rsic}$ parameters using the NYU Value-Added Galaxy Catalog (VAGC; \cite{blanton05}), which is based on the seventh release (DR7) of Sloan Digital Sky Survey (SDSS).
The stellar masses are estimated from the fits to the broad-band ($u$, $g$, $r$, $i$, and $z$ centered at 3540, 4770, 6230, 7630, and 9130 {\AA}) photometry (\cite{blanton07}) using a Chabrier IMF (\cite{chabrier03}) that we convert to the Salpeter IMF by adding a factor of 0.25 dex to them.
The $\rm{S\acute{e}rsic}$ parameters are measured in those five bands with a single component $\rm{S\acute{e}rsic}$ model fitting (\cite{blanton03}).
Considering completeness of the SDSS data (\cite{shen03}) and possible selection biases (\cite{franx08}), we use the VAGC sample between redshifts of 0.05 and 0.07, the $r$-band apparent magnitudes of 15.0 and 17.7 mag, and the $r$-band surface brightnesses brighter than 23 mag arcsec$^{-2}$, which contains a sample of $\sim$ $4\times10^{4}$ galaxies in total.
The criteria except for the redshift range affect weakly to the selection.
In practice, we have analyzed a sample selected with only the redshift criterion ($0.05\leq z \leq 0.07$), and found negligible change in the size distribution of the sample.
The median redshift of the SDSS galaxies is 0.062 at which the SDSS $z$-band corresponds to the rest-frame $\sim8600$ {\AA}.
We overplot in Figure \ref{fig:Ms_RecK} the effective radii measured in the SDSS $z$-band of the local galaxies as a function of stellar mass.
The late- and early-type galaxies are separated at the $\rm{S\acute{e}rsic}$ index in the $z$-band, $n$($z$), = 2.5.
We see for the both morphological types the LIRGs are comparable in size to or slightly ($\sim$ 20\% at a maximum) smaller than the local galaxies while the non-LIRGs are significantly (30--70\%) smaller than both the $z\sim1$ LIRGs and the local galaxies.
In particular, about a half of the low-$n(K_{\mathrm{s}})$ LIRGs have comparable sizes to the local disk-like galaxies.
In the local universe, a size dependence on the measured wavelength, namely a color gradient, between 11000 {\AA} and 8600 {\AA} is known to be only a few per cent (smaller in redder band) for both disk-like galaxies (\cite{deJong96}; \cite{barden05}) and early-type galaxies (\cite{mcintosh05}).
Even if we take into account this color gradient to estimate sizes of the local galaxies at $\lambda$ = 11000 {\AA}, our result would be unchanged.

\begin{figure}
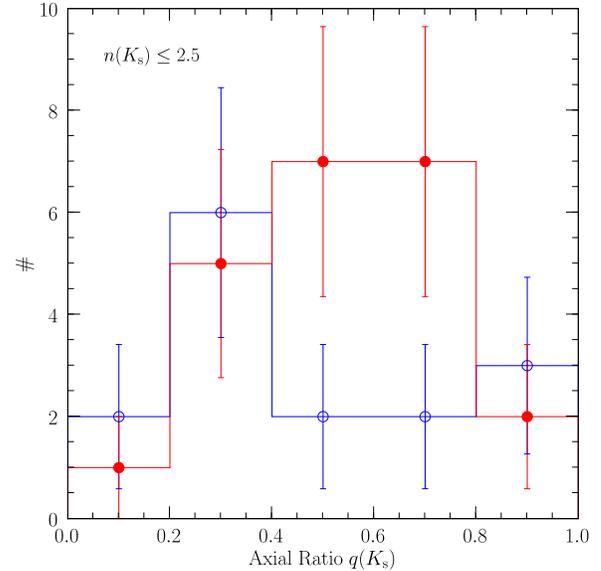

  \begin{center}
    \FigureFile(80mm,80mm){figure09.eps}
  \end{center}
  \caption{The histogram of axial ratio measured in the $K_{\mathrm{s}}$-band for the $z\sim1$ galaxies with $n$($K_{\mathrm{s}}$) $\leq2.5$ in the MODS. The symbols are the same as in Figure \ref{fig:Ms_Ks_morph}. Only the galaxies with $M_{\mathrm{s}}$= 3--15 $\times10^{10}$ $M_{\solar}$ are included, and the X-ray-detected galaxies are excluded. The error bars account for Poissonian errors only.}
  \label{fig:axrK_hist}
\end{figure}

\begin{figure}
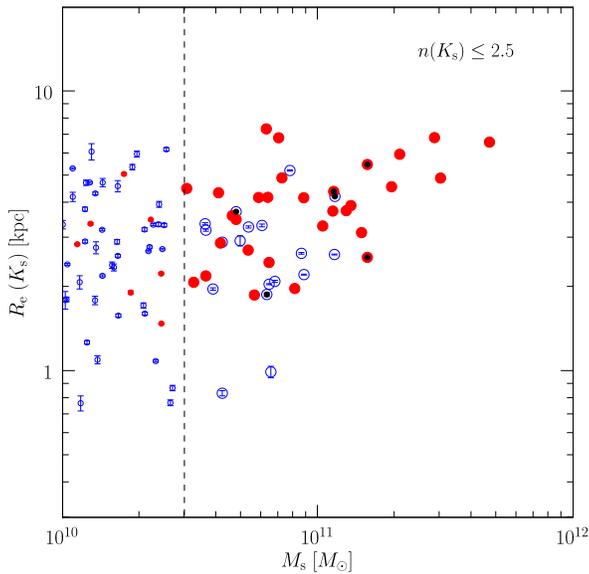

  \begin{center}
    \FigureFile(80mm,80mm){figure10.eps}
  \end{center}
  \caption{Similar to Figure \ref{fig:Ms_RecK}(a) but the sizes are described as the effective radii measured along to the semimajor axis. The symbols are the same as in Figure \ref{fig:Ms_Ks_morph}. The error bars indicate the uncertainties of the fitting. }
  \label{fig:Ms_ReK_disk}
\end{figure}

The trend that the non-LIRGs are further apart from the local relation compared to the LIRGs is in good agreement with a study by \citet{perez08b}, who used deeper MIPS 24 $\mu$m image than ours and found that 24$\mu$m-undetected ($f_{24}$ $<$ 15 $\mu$Jy) sources are smaller than other 24 $\mu$m sources for galaxies with $M_{\mathrm{s}}$ $>10^{11}$ $M_{\solar}$.
We find that the larger difference in circularized size between the low-$n(K_{\mathrm{s}})$ LIRGs and low-$n(K_{\mathrm{s}})$ non-LIRGs is partly caused by the fact that those two groups have different distributions of axial ratio, which is shown in Figure \ref{fig:axrK_hist}.
For a comparison of axial ratio, here we limit the sample within $M_{\mathrm{s}}$ = 3--15 $\times10^{10}$ $M_{\solar}$ which is the stellar mass range of the low-$n(K_{\mathrm{s}})$ non-LIRGs.
A median axial ratio and a scatter for the LIRGs and non-LIRGs are $0.57\pm0.21$ and $0.38\pm0.26$, respectively.
Taking into account the difference in axial ratio, we replot the stellar mass-size relation at $z\sim1$ using not the \textit{circularized} sizes ($R_{\mathrm{e,c}}$) but the semimajor axis sizes ($R_{\mathrm{e}}$) in Figure \ref{fig:Ms_ReK_disk}.
The LIRGs are still larger than the non-LIRGs although the size difference between the LIRGs and non-LIRGs seen in Figure \ref{fig:Ms_RecK} becomes small.

Although some of the low-$n(K_{\mathrm{s}})$ non-LIRGs with low axial ratio ($q[K_{\mathrm{s}}]<0.4$) show redder $U-V$ colors than the low-$n(K_{\mathrm{s}})$ LIRGs which may be caused by their high inclination angles, we cannot investigate further the cause of the different $q(K_{\mathrm{s}})$ distributions considering the small sample size.

%%%%%%%%%%%%%%%%%%%%%%%%%%%%%%%%%%%%%%%%%%%%%%%%%%%%%%%%%%%%%%%%%%%%%%%%%%%%%%%
\subsection{Concentration-Asymmetry indices}
\label{sect:CA_K}

\begin{figure}
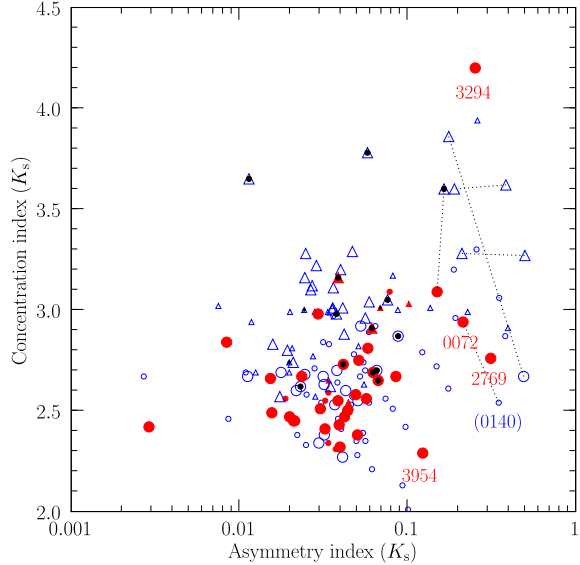

  \begin{center}
    \FigureFile(80mm,80mm){figure11.eps}
  \end{center}
  \caption{Asymmetry ($A$) and Concentration ($C$) indices of the $z\sim1$ galaxies measured in the MODS deep $K_{\mathrm{s}}$-band image. The symbols are the same as in Figure \ref{fig:Ms_Ks_morph}. The \textit{dotted} lines connect probable physically-interacting pairs in the massive ($M_{\mathrm{s}}$ $\geq3\times10^{10}$ $M_{\solar}$) sample where two galaxies have close spectroscopic redshifts as well as being located closely ($\leq$ 1.5 arcsec) in the $K_{\mathrm{s}}$-band image. The galaxy ID:0072 has a companion (ID:0140) which is in the less massive ($M_{\mathrm{s}}=$ 1--3 $\times10^{10}$ $M_{\solar}$) sample. The galaxy ID:2769 ($z_{\mathrm{phot}}$ = 0.98) shows a high $A$ due to the existence of a companion (ID:2722) which have a spectroscopic redshift of 0.581 (not shown in the figure); This pair is just due to a random projection. The galaxy ID:3294 is located close to very bright source. The galaxy ID:3954 is located close to the edge of the image.}
  \label{fig:asymK_concK}
\end{figure}

\begin{figure}
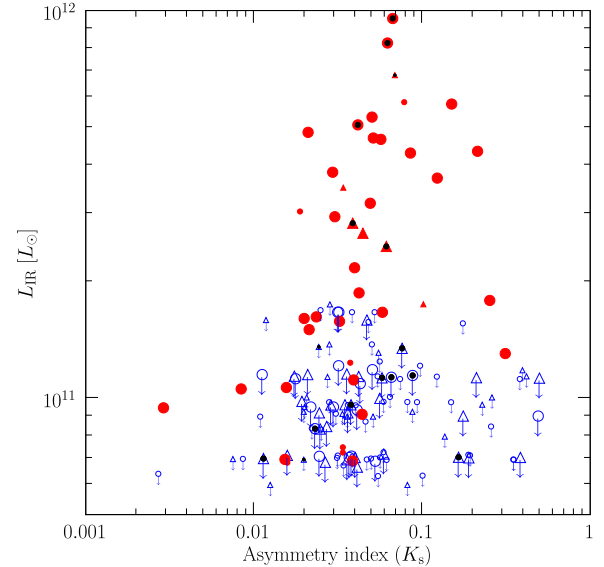

  \begin{center}
    \FigureFile(80mm,80mm){figure12.eps}
  \end{center}
  \caption{IR luminosities of the $z\sim1$ galaxies in the MODS as a function of Asymmetry index measured in the $K_{\mathrm{s}}$-band. The symbols are the same as in Figure \ref{fig:Ms_Ks_morph}. The non-LIRGs (i.e., 24 $\mu$m-undetected galaxies) are also marked with down-pointing arrows as their upper limits of $L_{\mathrm{IR}}$ corresponding to $f_{24}$ $=80$ $\mu$Jy at their redshifts.}
  \label{fig:asymK_LIR}
\end{figure}

Figure \ref{fig:asymK_concK} shows the result of Concentration-Asymmetry ($C$-$A$) measurement in the MODS deep $K_{\mathrm{s}}$-band image of the $z\sim1$ galaxies.
Since the wavelength in which the indices measured is longer than that in previous (optical) studies, separation criteria of morphological type of galaxies on the $C$-$A$ plane established by those studies (e.g., \cite{conselice03}) are not applicable on our (rest-frame $J$-band) $C$-$A$ plane.
Therefore, it is difficult to interpret absolute values of $C$ and $A$ measured in the $K_{\mathrm{s}}$-band.
Nonetheless, relative comparison of those values within the $z\sim1$ sample (LIRGs versus non-LIRGs or low-$n(K_{\mathrm{s}})$ versus high-$n(K_{\mathrm{s}})$ sample) is meaningful to investigate the difference between the populations.
Figure \ref{fig:asymK_concK} shows the consistent distribution of morphologies with the result of the \textsc{Galfit} analysis.
Most of the LIRGs have low $C(K_{\mathrm{s}})$ compared to the non-LIRGs which are distributed widely in $C(K_{\mathrm{s}})$.
That is the similar trend as seen in the classification with $\rm{S\acute{e}rsic}$ index.
We actually find that $C(K_{\mathrm{s}})$ correlates with $n(K_{\mathrm{s}})$ well in our sample.

A part of galaxies show larger $A(K_{\mathrm{s}})$ than the others. 
At the index $A(K_{\mathrm{s}})>0.1$, there are 12 galaxies ($\sim15$\% of the total) for the massive sample.
High $A(K_{\mathrm{s}})$ indices of two out of 12 galaxies are partly affected by nearby bright source (ID:3294) or by located near the edge of the image (ID:3954).
The remaining 10 galaxies have a companion very close ($\leq$ 1.5 arcsec or $\leq$ 12 kpc in projected distances) to themselves.
Among the 5 pairs, 4 pairs have almost the same spectroscopic redshift to each other, which means that they can be physically interacting, not a random projection.
Those interacting pairs are connected with \textit{dotted} lines in Figure \ref{fig:asymK_concK}.
The remaining one pair might be just due to the projection; the galaxy ID:2769 have a photometric redshift of 0.98 while the companion have a spectroscopic redshift of 0.581.
Therefore, highly asymmetric features observed in the $z\sim1$ sample are thought to be induced mainly by strong interaction (i.e., major-merging events).
IR luminosities (an indicator of dusty star formation activity) are shown as a function of the asymmetry indices in Figure \ref{fig:asymK_LIR}.
The median and 1-$\sigma$ scatter of the index $A(K_{\mathrm{s}})$ for the LIRGs and the non-LIRGs are 0.039 $\pm$ 0.076 and 0.034 $\pm$ 0.105, respectively.
We find no correlation between star formation and NIR asymmetry property.

%%%%%%%%%%%%%%%%%%%%%%%%%%%%%%%%%%%%%%%%%%%%%%%%%%%%%%%%%%%%%%%%%%%%%%%%%%%%%%%
\section{Discussion}
\label{sect:discussion}

As seen in Section \ref{sect:mips}, our sample contains majority of $z\sim1$ LIRGs which are thought to be a major star-forming population at that epoch.
Quantitative morphological analysis reveals that those LIRGs show low-$n(K_{\mathrm{s}})$ (disk-like) structure in the NIR.
This is consistent with the result of the AO-supported study in the $K$-band of \citet{mel08}.
Also, the low-$n(K_{\mathrm{s}})$ LIRGs consist of more than a half of the whole low-$n(K_{\mathrm{s}})$ galaxies at $z\sim1$ above $M_{\mathrm{s}}$ $\geq$ 3 $\times 10^{10}$ $M_{\solar}$.
Those observational evidences indicate that the star formation at $z\sim1$ is mainly occurred in relatively massive disk-like galaxies, and that those star-forming massive disk-like galaxies are popular at $z\sim1$.
In addition, the fact that the low-$n(K_{\mathrm{s}})$ galaxies in our sample does not show any correlation between the asymmetry and the IR luminosity except for a few close pairs indicates that the star formation is not triggered mainly by early-stage major-merging events which disturb the NIR morphology strongly while it is still possible that some of the LIRGs are in late stage of interaction.
As a next step, we investigate rest-frame UV-to-optical morphologies of the low-$n(K_{\mathrm{s}})$ galaxies to understand how the star formation is occurred in the massive disk-like galaxies.

We use the ACS $V_{606}$- and $z_{850}$-bands (rest-frame $U$- and $B$-bands) images and the source catalogs version 2.0 provided by the GOODS project to examine the distribution of star-forming regions in the $z\sim1$ galaxies, and to compare them with those of local disk-like galaxies.
First, we find a $z_{850}$-band counterpart for all the $z\sim1$ ($K_{\mathrm{s}}$-selected) galaxies.
For galaxies with multiple $z_{850}$-band counterparts, we select the closest one as a counterpart.
Two of the 139 $z\sim1$ galaxies (ID=3294, 3844) have no $z_{850}$-band counterparts.
Both of them are close to a bright star, which may have made the $z_{850}$-band source detection failed in those region.
We do not try to tune detection parameters to detect them, but remove them from the current discussion.
Then, we perform the profile fitting for the $z_{850}$-band counterparts using \textsc{Galfit} in a similar manner as for the $K_{\mathrm{s}}$-band galaxies except for fitting the background sky level simultaneously in the $z_{850}$-band image.
Similar to the $K_{\mathrm{s}}$-band morphological analysis, we use the \textsc{SExtractor} outputs as initial guesses except for $n$ which is set to 1.5.
We use a weight map produced by the GOODS team for the signal-to-noise per pixel.
We create a PSF image for each ACS tile by stacking spectroscopically-identified, unsaturated stars in the tile.
We confirm that even if we smooth and resample the $z_{850}$-band image to match the image qualities (i.e., pixel scale and PSF size) to those of the $K_{\mathrm{s}}$-band image, the fitting results are little changed in a statistical sense in comparison with results evaluated with the original $z_{850}$-band image.
Then, we evaluate the $V_{606}$-band morphologies of the galaxies in the similar manner.
Among the $z_{850}$-band structural parameters, the semimajor, semiminor axial radius, the axial ratio, and the position angle are used as initial guesses for \textsc{Galfit}.
The position $x$, $y$ are fixed at those in the $z_{850}$-band.
The initial guess of the total magnitude is obtained from the GOODS-N public catalog.

\begin{figure}
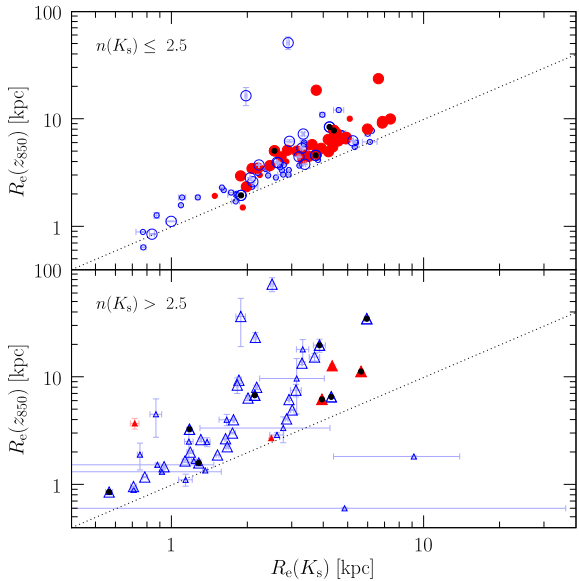

  \begin{center}
    \FigureFile(80mm,80mm){figure13.eps}
  \end{center}
  \caption{The comparison of effective radii measured in the $K_{\mathrm{s}}$- and $z_{850}$-band. The symbols are the same as in Figure \ref{fig:Ms_Ks_morph}. The dotted diagonal indicates where the parameters derived in the $z_{850}$-band are equal to those in the $K_{\mathrm{s}}$-band. The error bars indicate the uncertainties of the fitting.}
  \label{fig:reK_rezorg}
\end{figure}

Figure \ref{fig:reK_rezorg} shows the comparison of the effective radii along the semimajor axis measured in the $K_{\mathrm{s}}$-band and $z_{850}$-band.
We mention that there is no significant difference in $\rm{S\acute{e}rsic}$ index and axial ratio between the rest-frame UV-to-optical and NIR bands.
The $R_{\mathrm{e}}$($K_{\mathrm{s}}$) is systematically smaller than the $R_{\mathrm{e}}$($z_{850}$) irrespective of the IR luminosity; a mean size and an uncertainty of the mean of $R_{\mathrm{e}}$($K_{\mathrm{s}}$)/$R_{\mathrm{e}}$($z_{850}$) $\sim$ $0.65\pm0.01$ and $0.44\pm0.01$ for the low- and high-$n(K_{\mathrm{s}})$ massive sample, respectively.
The difference between rest-frame $J$-band and $B$-band sizes found in our $z\sim1$ sample is a factor of 3--5 larger than that for local galaxies.
Similar trend of larger color gradient of $z\sim1$ galaxies was reported by \citet{trujillo07}.
They used the ACS F814W and Near-Infrared Camera and Multi-Object Spectrometer (NICMOS) F160W for the size comparison of 27 galaxies with $M_{\mathrm{s}}$ $>10^{11}$ $M_{\solar}$ at $0.8<z<1.8$, and found that sizes measured in F160W are $19\pm7$\% smaller than those in F814W.
Previous optical studies which investigated sizes of disk-like galaxies with $M_{\mathrm{s}}$ = $10^{10}$--$10^{11}$ $M_{\solar}$ in the rest-frame optical band report that there is no size evolution from $z\sim1$ to the present (\cite{barden05}; \cite{trujillo07}).
Our data confirms their result; the circularized sizes of the $z\sim1$ low-$n(K_{\mathrm{s}})$ sample measured in the ACS $z_{850}$-band (rest-frame 4250 {\AA}) are comparable with those of the local galaxies measured in the SDSS $g$-band (rest-frame 4360 {\AA}) in the same stellar mass range.
Combined with the systematic size difference seen in the rest-frame $J$-band (Figure \ref{fig:Ms_RecK}), the $z\sim1$ disk-like galaxies seem to have a steeper color gradient from the local disk-like galaxies.
\begin{figure}
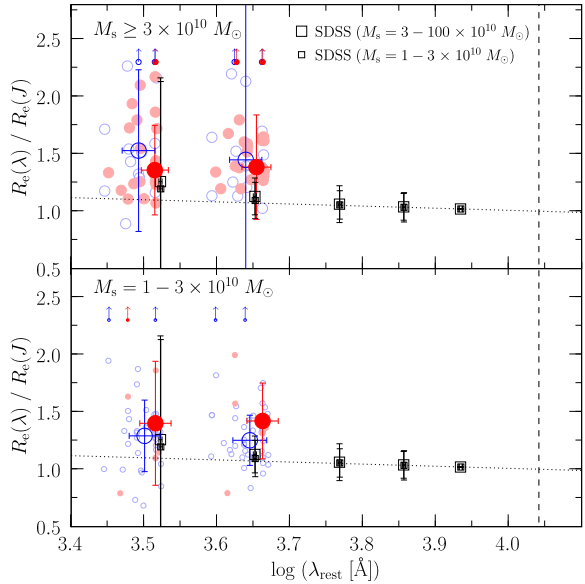

  \begin{center}
    \FigureFile(80mm,80mm){figure14.eps}
  \end{center}
  \caption{The normalized effective radii as a function of rest-frame wavelength of the $z\sim1$ disk-like galaxies. Sizes are normalized at $J$-band (\textit{dashed} line). The symbols except for those with error bars and black dots are the same as in Figure \ref{fig:Ms_Ks_morph}. The large circles with error bar represent the medians and scatters. Two stellar mass ranges ($M_{\mathrm{s}}$ $\geq3\times10^{10}$ $M_{\solar}$, $M_{\mathrm{s}}$= 1--3 $\times10^{10}$ $M_{\solar}$) are shown separately. All X-ray-detected galaxies are excluded. Galaxies whose normalized radius is beyond the vertical axis scale are plotted with an up-pointing arrow. The \textit{squares} with error bar show the median and 1-$\sigma$ scatter of the radii for local disk-like galaxies used in Section \ref{sect:Ms_Re_K}. The \textit{dotted} line represents the color gradient (a slope $\alpha=-0.184$) for local disk-like galaxies (\cite{barden05}; \cite{deJong96}) where we set the intercept to make the size at $J$-band equal to unity.}
  \label{fig:color_grad}
\end{figure}
We show the comparison of the color gradient of disk-like galaxies at $z\sim1$ and $z=0$ in Figure \ref{fig:color_grad}.
For the $z\sim1$ galaxies with $n$($K_{\mathrm{s}}$) $\leq2.5$, the sizes in the $V_{606}$- and $z_{850}$-band normalized at the $K_{\mathrm{s}}$-band are plotted.
The \textit{dotted} line is an observed color gradient of local disk-like galaxies (\cite{barden05}) where the slope, $\alpha=-0.184$, is obtained using the VAGC $\rm{S\acute{e}rsic}$ parameters in a similar manner to us.
The intercept of the \textit{dotted} line is determined to make the size at $J$-band equal to unity.
\textit{Open squares} show the relative sizes of the local disk-like ($n$[$z$] $\leq2.5$) galaxies in our SDSS/VAGC catalog (used in Section \ref{sect:Ms_Re_K}) normalized at $z$-band and shifted vertically to make the size at $z$-band lie on the \textit{dotted} line.
We find a steeper slope of the color gradient at $z\sim1$ than at $z=0$ by a linear fit with the intercept fixed to unity at the $J$-band.
The slope at $z\sim1$ derived using the ACS $V_{606}$, $z_{850}$, and the MOIRCS $K_{\mathrm{s}}$-band sizes is $\alpha=-0.77 \pm 0.10$ ($-0.97 \pm 0.12$) and $-0.98 \pm 0.03$ ($-0.56 \pm 0.03$) for the LIRGs and the non-LIRGs in the massive (less massive) sample, respectively.
On the contrary, the slope at $z=0$ derived from $u$, $g$, and $z$-band sizes of our VAGC catalog is $\alpha=-0.19 \pm 0.00$ for the both stellar mass ranges.
We note that the linear fit with all the five SDSS bands yields the same slope of -0.19 as with the above three bands.
Consequently, the $z\sim1$ galaxies with $M_{\mathrm{s}}$ $\geq 10^{10}$ $M_{\solar}$ have 3--5 times steeper color gradient than the local galaxies at the similar stellar mass range.
The change in the color gradient between $z\sim1$ and $z=0$ can be interpreted as that while the distribution of young stellar population indicated by the rest-frame UV--optical sizes is unchanged from $z\sim1$ to the present, the structure consisting of old stellar population indicated by the rest-frame NIR sizes is still being constructed at $z\sim1$.
Additionally, changes in dust distribution in a galaxy should also contribute to the decrease of the color gradient.

Recent NIR Integral Field Unit (IFU) observations investigating the kinematics of distant ($z \gtrsim 0.6$) disk galaxies revealed that they show a similar Tully-Fisher relation to the local disk galaxies (\cite{flores06}; \cite{puech08}; \cite{lemoine10}).
In addition, \citet{neichel08} investigated a radial distribution of color and SFR along disk of rotating spiral disk galaxies which show similar morphology and kinematics to the local disk galaxies, and found that the very active and recent star formation  is occurred in the outer parts of the disks.
They concluded the observational evidence as a rapid, inside-out disk formation of disk galaxies.
Our results are consistent with such a disk formation scenario, in which local disk galaxies form their disk structure from the center to the outskirt.
Intense star formation activity observed as LIRGs at $z\sim1$ may be mostly related to the formation and growth of disk structure in relatively massive galaxies, and contribute significantly to the high cosmic SFR density at $z\sim1$.

%%%%%%%%%%%%%%%%%%%%%%%%%%%%%%%%%%%%%%%%%%%%%%%%%%%%%%%%%%%%%%%%%%%%%%%%%%%%%%%
\section{Summary}
\label{sect:summary}

In this paper, we studied rest-frame NIR morphologies of galaxies at $z=$ 0.8--1.2 from the deep NIR imaging survey (MODS; K09) catalog in a part of the GOODS-N field.
The $K_{\mathrm{s}}$-band image, which is the key data for this work, covers $\sim$ 28 arcmin$^2$ and has a PSF $\sim$ \timeform{0''.4} (FWHM) which corresponds to 3.2 kpc at $z\sim1$.
Cross-correlating our $K_{\mathrm{s}}$-band galaxy catalog with the \textit{Spitzer}/MIPS 24 $\mu$m and the \textit{Chandra} X-ray source catalogs, we identified LIRGs and AGN candidates.
Using \textsc{Galfit}, we performed a two-dimensional light profile fitting of the $z\sim1$ galaxies in the $K_{\mathrm{s}}$-band with a single component $\rm{S\acute{e}rsic}$ model.
Our simulation with artificial galaxy images showed that our morphological analysis is reliable up to $K_{\mathrm{s}}$(AB) $\sim$ 22.5 mag irrespective of morphology.
That magnitude limit corresponds to $M_{\mathrm{s}}$ $\sim1\times10^{10}$ $M_{\solar}$ at $z\sim1$ and we adopted that stellar mass as a selection criterion.
As a non-parametric morphological study, we measured the Concentration ($C$) and Asymmetry ($A$) indices of the galaxies in the $K_{\mathrm{s}}$-band.

We investigated NIR morphological properties of the 139 galaxies having $z=$ 0.8--1.2 and $M_{\mathrm{s}}$ $\geq1\times10^{10}$ $M_{\solar}$.
The main results about those galaxies are summarized as follows.

\begin{itemize}
 \item $\sim90$\% of LIRGs show disk-like ($n$[$K_{\mathrm{s}}$] $\sim$ 1--2) light profiles in the $K_{\mathrm{s}}$-band.
 \item Those low-$n(K_{\mathrm{s}})$ LIRGs consist of 60\% of the whole low-$n(K_{\mathrm{s}})$ sample above $M_{\mathrm{s}}$ $\geq$ 3 $\times 10^{10}$ $M_{\solar}$.
 \item About a half of the low-$n(K_{\mathrm{s}})$ LIRGs are comparable in size and the others are slightly ($\sim$ 20\% at a maximum) small compared to local disk-like galaxies.
 \item No correlation between the NIR asymmetry properties and $L_{\mathrm{IR}}$ (star formation activity) is found.
\end{itemize}

Those results indicate that the star formation at $z\sim1$ is (i) mainly occurred in most relatively massive disk galaxies, and (ii) is not triggered by early-phase strong galaxy-galaxy interactions which disturb strongly the NIR morphology.

In order to investigate how the star formation is occurred in the massive disk-like galaxies, we compared the rest-frame $J$-band morphologies of the $z\sim1$ galaxies with the rest-frame $U$- and $B$-band ones using the \textit{HST}/ACS $V_{606}$- and $z_{850}$-band image, respectively.
There is no significant difference in $\rm{S\acute{e}rsic}$ index and axial ratio between the rest-frame UV--optical and NIR bands.
The comparison of the effective radii shows a $\sim30$\% systematic offset where sizes in redder band are smaller.
Although this is known as ``color gradients'' of galaxies in the local universe, the gradient we found is 3--5 times steeper than in the local universe.
Since we found the fact that the rest-frame optical sizes of the $z\sim1$ galaxies are comparable to the local galaxies, this steeper color gradient indicates that fundamental disk structure in those massive galaxies is still being constructed at $z\sim1$.
Our results indicate not only that more than a half of relatively massive disk-like galaxies at $z\sim1$ are in violent star formation epochs observed as LIRGs but also that most of those LIRGs are constructing their disk structure vigorously.
The high star formation rate density in the universe at $z\sim1$ may be dominated by star formation in disk region in massive galaxies.

%%%%%%%%%%%%%%%%%%%%%%%%%%%%%%%%%%%%%%%%%%%%%%%%%%%%%%%%%%%%%%%%%%%%%%%%%%%%%%%
\bigskip

We would like to thank the Subaru Telescope staff for their invaluable help and support for observations.
We are grateful to an anonymous referee for many helpful comments and suggestions, that improved our paper.
This study is based on data collected at Subaru Telescope, which is operated by the National Astronomical Observatory of Japan.
This work is based on in part on observations made with the \textit{Spitzer Space Telescope}, which is operated by the Jet Propulsion Laboratory, California Institute of Technology under a contract with NASA.
The Image Reduction and Analysis Facility (IRAF) used in this paper is distributed by the National Optical Astronomy Observatories, which are operated by the Association of Universities for Research in Astronomy, Inc., under cooperative agreement with the National Science Foundation.
The SDSS Web Site is http://www.sdss.org/.
The SDSS is managed by the Astrophysical Research Consortium for the Participating Institutions. The Participating Institutions are the American Museum of Natural History, Astrophysical Institute Potsdam, University of Basel, University of Cambridge, Case Western Reserve University, University of Chicago, Drexel University, Fermilab, the Institute for Advanced Study, the Japan Participation Group, Johns Hopkins University, the Joint Institute for Nuclear Astrophysics, the Kavli Institute for Particle Astrophysics and Cosmology, the Korean Scientist Group, the Chinese Academy of Sciences (LAMOST), Los Alamos National Laboratory, the Max-Planck-Institute for Astronomy (MPIA), the Max-Planck-Institute for Astrophysics (MPA), New Mexico State University, Ohio State University, University of Pittsburgh, University of Portsmouth, Princeton University, the United States Naval Observatory, and the University of Washington.

%%%%%%%%%%%%%%%%%%%%%%%%%%%%%%%%%%%%%%%%%%%%%%%%%%%%%%%%%%%%%%%%%%%%%%%%%%%%%%%
\appendix

%%%%%%%%%%%%%%%%%%%%%%%%%%%%%%%%%%%%%%%%%%%%%%%%%%%%%%%%%%%%%%%%%%%%%%%%%%%%%%%
\section{Properties of the MODS $z\sim1$ galaxies}
\label{sect:GT2_catalog}

Table \ref{tab:GT2morph_catalog_sample} is a sample of the information about the properties of the $z\sim1$ galaxies analyzed in this paper.
Columns 1--6 list the basic properties of the galaxy derived from Kajisawa et al. (2009, K09).
Column 1 indicates the galaxy identification number.
Columns 2 and 3 are the Right Ascension (R.A.) and Declination (Dec.) of the galaxy for epoch J2000 in degrees.
Column 4 lists the apparent $K_{\mathrm{s}}$-band magnitude in the AB system.
Column 5 is the measured redshift of the galaxy.
Column 6 specifies whether the redshift was determined spectroscopically (s) or photometrically (p).
Column 7 is the stellar mass of the galaxy in units of $10^{10} M_{\solar}$.
Column 8 is the MIPS 24 $\mu$m flux in units of $\mu$Jy derived from the 24 $\mu$m source catalog (M. Dickinson et al. in preparation; R. Chary et al. in preparation).
Possible 24 $\mu$m sources which would be caused by confusion effect identified by visual inspection on the MIPS 24 $\mu$m image are denoted by an asterisk.
Column 9 is the IR luminosity in units of $10^{11} L_{\solar}$.
Columns 10--14 list the morphological properties of the galaxy measured in the $K_{\mathrm{s}}$-band.
Column 10 indicates the value of the S$\acute{\mathrm{e}}$rsic index of the fit to the galaxy.
Column 11 is the effective radius along the semimajor axis of the galaxy.
Column 12 is the axial ratio of the galaxy.
And finally, Column 13 and 14 are the concentration and asymmetry indices of the galaxy.
Galaxies listed in the X-ray catalog (\cite{alex03}) are denoted as ``X-ray'' in the Comment.
The entire catalog for all the 155 galaxies including the X-ray-detected galaxies analyzed in this paper is available as an online material.

\begin{table*}
  \begin{center}
  \caption{Properties of the MODS $z\sim1$ galaxies.}
  \label{tab:GT2morph_catalog_sample}
\scalebox{0.8}{%
    \begin{tabular}{ccccllcccccccc|c}
\hline \hline
ID & R.A. & Dec. & $K_{\rm{s}}$(AB) & \multicolumn{2}{c}{Redshift} & $M_{\rm{s}}
$ & $f_{24}$ & $L_{\rm{IR}}$ & $n$($K_{\rm{s}}$) & $R_{\rm{e}}$($K_{\rm{s}}$) & 
$q$($K_{\rm{s}}$) & $C(K_{\rm{s}})$ & $A(K_{\rm{s}})$ & Comment \\
& [deg] & [deg] & [mag] &  &  &  [$10^{10} M_{\solar}$] & [$\mu$Jy] & [$10^{11} L
_{\solar}$] &  & [kpc] &  &  & &  \\
(1) & (2) & (3) & (4) & (5) & (6) & (7) & (8) & (9) & (10) & (11) & (12) & (13) 
& (14) \\ \hline
0072 & 189.074141 & +62.235512 &  20.51 & 0.846 & (s) &  20.9 & 427.0 &  4.3 &  2.2 &   6.0 & 0.75 & 2.94 & 0.214 &   \\
0132 & 189.159586 & +62.197463 &  20.34 & 0.841 & (s) &  10.4 & 230.0 &  2.2 &  0.9 &   3.3 & 0.55 & 2.32 & 0.040 &   \\
0151 & 189.194153 & +62.180282 &  20.66 & 0.940 & (s) &   5.6 & 354.0 &  4.7 &  1.1 &   1.9 & 0.60 & 2.75 & 0.051 &   \\
0152 & 189.157521 & +62.197045 &  20.32 & 0.839 & (s) &   7.0 & 173.0 &  1.6 &  0.7 &   6.8 & 0.20 & 2.41 & 0.032 &   \\
0156 & 189.153342 & +62.203661 &  18.81 & 0.848 & (s) &  47.0 & 379.0 &  3.8 &  2.0 &   6.6 & 0.85 & 2.98 & 0.029 &   \\
0167 & 189.172377 & +62.191551 &  19.86 &  0.99 & (p) &  19.4 & 287.0 &  4.3 &  1.2 &   4.6 & 0.80 & 2.67 & 0.085 &   \\
0227 & 189.169444 & +62.193243 &  21.35 &  0.99 & (p) &   4.2 & 110.0 &  1.5 &  1.0 &   2.9 & 0.37 & 2.45 & 0.021 &   \\
0265 & 189.131821 & +62.212030 &  21.28 & 1.016 & (s) &   2.2 & 109.0 &  1.6 &  1.2 &   3.5 & 0.36 & 2.55 & 0.032 &   \\
0510 & 189.179991 & +62.196678 &  20.20 & 1.007 & (s) &  11.4 & 113.0 &  1.6 &  1.7 &   3.7 & 0.66 & 2.67 & 0.024 &   \\
0658 & 189.192432 & +62.195025 &  19.36 & 1.016 & (s) &  28.6 & 290.0 &  4.7 &  1.3 &   6.8 & 0.73 & 2.56 & 0.057 &   \\
\hline
    \end{tabular}}
  \end{center}
\end{table*}

%%%%%%%%%%%%%%%%%%%%%%%%%%%%%%%%%%%%%%%%%%%%%%%%%%%%%%%%%%%%%%%%%%%%%%%%%%%%%%%
\section{Images of the MODS $z\sim1$ galaxies}

\begin{figure*}
  \begin{center}
   \includegraphics[width=16.0cm,clip]{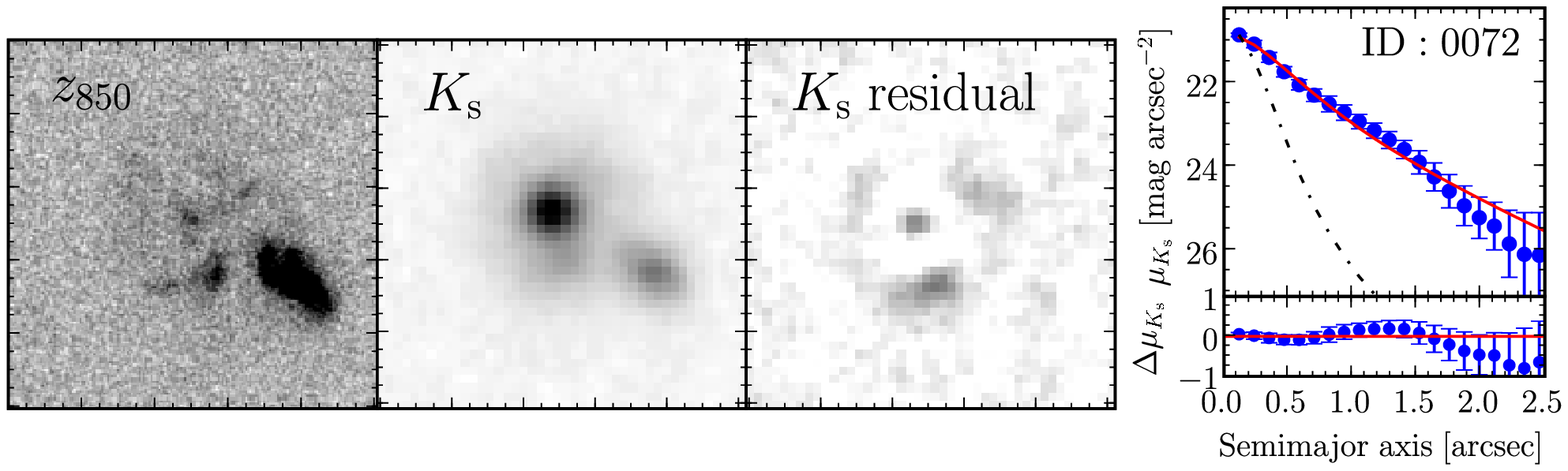} \\
   \includegraphics[width=16.0cm,clip]{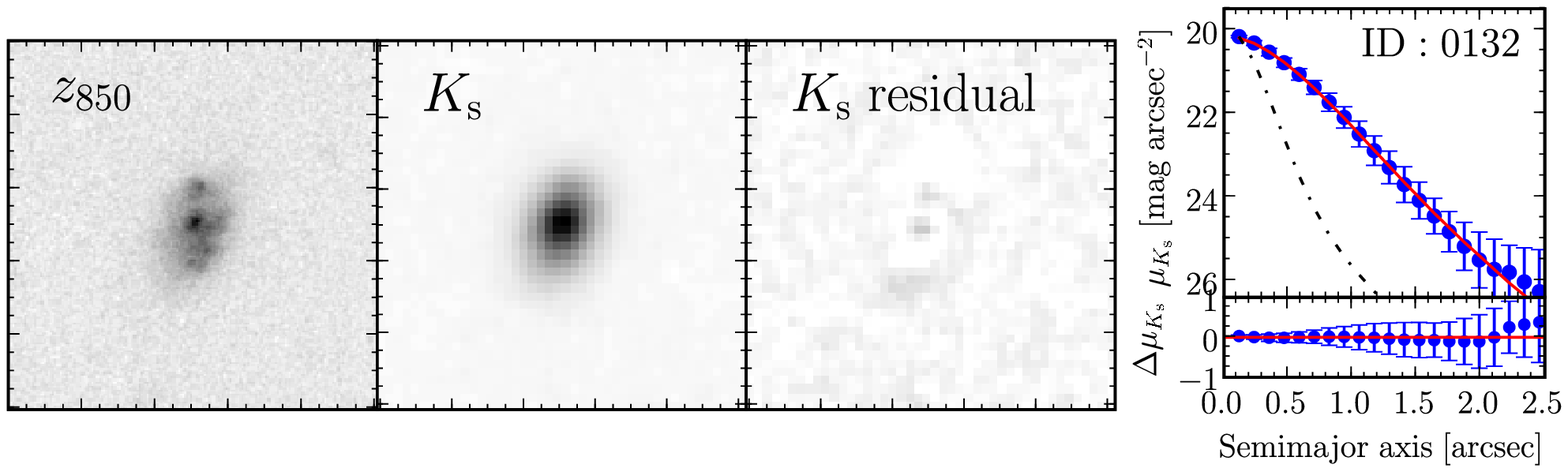} \\
  \end{center}
  \caption{LIRGs with $n(K_{\mathrm{s}})$ $\leq2.5$ at $z\sim1$ in the MODS deep region. The first two galaxies are shown as a sample. Iimage lists for all the 155 galaxies including the X-ray-detected galaxies analyzed in this paper are available as online materials. \textit{Left panels}: ACS $z_{850}$-band (rest-frame $B$), $K_{\mathrm{s}}$-band (rest-frame $J$) images, and $K_{\mathrm{s}}$-band \textsc{galfit} residual image after model subtraction from the $K_{\mathrm{s}}$-band image are shown from \textit{left} to \textit{right}. North is to the top and the East is to the left. The size of each image is approximately 5.0 $\times$ 5.0 arcsec ($\sim40\times40$ kpc). The residual image is shown with narrower dynamic range than that for the $K_{\mathrm{s}}$-band image to display the residual pattern clearly. Simultaneously-fitted neighbors are subtracted from the residual image as well with their best-fit S$\acute{\mathrm{e}}$rsic models. \textit{Right panels}: Observed (filled \textit{circles}) and best-fit (\textit{solid line}) surface brightness profiles in the $K_{\mathrm{s}}$-band along the semimajor axis are shown in the \textit{top}. The profiles are derived with elliptical isophote fitting package \texttt{ellipse} in IRAF. The shape (axial ratio and position angle) and the center of the ellipse are fixed with values derived using \textsc{galfit}, and the isophote radius is changed from 0.0 to 2.5 arcsec. The observed profile is derived from the $K_{\mathrm{s}}$-band image subtracted the simultaneously-fitted neighbors with their best-fit models, and is plotted until the radius reaches 1.5 times the semimajor axial length of the Kron elliptical aperture, which is a criterion for choosing neighbors, or 2.5 arcsec at a maximum. Error bars represent 1-$\sigma$ scatter of intensity data along the ellipse at a given radius. The \textit{dot-dashed} line shows the profile of the PSF used in \textsc{galfit} normalized to match the brightness with the observed point at the center. A residual of the observed profile is shown in the \textit{bottom}.}
  \label{fig:GT2morph_stamp_sample}
 \end{figure*}

Figure \ref{fig:GT2morph_stamp_sample} is a sample of postage stamp images and $K_{\mathrm{s}}$-band surface brightness profiles of the $z\sim1$ sample analyzed in this paper.
In the \textit{left panels}, ACS $z_{850}$-band (rest-frame $B$), $K_{\mathrm{s}}$-band (rest-frame $J$) images, and $K_{\mathrm{s}}$-band \textsc{galfit} residual image after model subtraction from the $K_{\mathrm{s}}$-band image are shown from \textit{left} to \textit{right}. 
North is to the top and the East is to the left. 
The size of each image is approximately 5.0 $\times$ 5.0 arcsec ($\sim40\times40$ kpc). 
The residual image is shown with narrower dynamic range than that for the $K_{\mathrm{s}}$-band image to display the residual pattern clearly. 
Simultaneously-fitted neighbors are subtracted from the residual image as well with their best-fit S$\acute{\mathrm{e}}$rsic models. 
In the \textit{right panels}, observed (filled \textit{circles}) and best-fit (\textit{solid line}) surface brightness profiles in the $K_{\mathrm{s}}$-band along the semimajor axis are shown in the \textit{top}. 
The profiles are derived with elliptical isophote fitting package \texttt{ellipse} in IRAF. 
The shape (axial ratio and position angle) and the center of the ellipse are fixed with values derived using \textsc{galfit}, and the isophote radius is changed from 0.0 to 2.5 arcsec. 
The observed profile is derived from the $K_{\mathrm{s}}$-band image subtracted the simultaneously-fitted neighbors with their best-fit models, and is plotted until the radius reaches 1.5 times the semimajor axial length of the Kron elliptical aperture, which is a criterion for choosing neighbors, or 2.5 arcsec at a maximum. 
Error bars represent 1-$\sigma$ scatter of intensity data along the ellipse at a given radius. 
The \textit{dot-dashed} line shows the profile of the PSF used in \textsc{galfit} normalized to match the brightness with the observed point at the center. 
A residual of the observed profile is shown in the \textit{bottom}.
Image lists for all the 155 galaxies including the X-ray-detected galaxies analyzed in this paper are available as online materials.

%%%%%%%%%%%%%%%%%%%%%%%%%%%%%%%%%%%%%%%%%%%%%%%%%%%%%%%%%%%%%%%%%%%%%%%%%%%%%%%

\end{document}